\documentclass[11pt]{amsart} 
\usepackage{graphicx}
\usepackage{amssymb, amsmath}
\usepackage{amsfonts}
\usepackage{amssymb}
\usepackage{float}

\newtheorem{theorem}{Theorem}

\newtheorem{corollary}[theorem]{Corollary}

\newtheorem{definition}[theorem]{Definition}

\newtheorem{remark}[theorem]{Remark}

\newcommand{\divv}{\text{\rm div}}
\newcommand{\ind}{\text{\rm ind}}

\newcommand{\cL}{\mathcal L}
\newcommand{\C}{\mathbb C}
\newcommand{\N}{\mathbb N}

\newcommand{\R}{\mathbb R}

\newcommand{\dist}{\text{\rm dist}}
\numberwithin{equation}{section}
\numberwithin{theorem}{section}
\numberwithin{example}{section}

\numberwithin{figure}{section}

\begin{document}

\title[Separations in Taylor-Couette-Poiseuille Flow]{Boundary-Layer and Interior Separations in  the Taylor--Couette--Poiseuille Flow}

\author[Ma]{Tian Ma}
\address[TM]{Department of Mathematics, Sichuan University,
Chengdu, P. R. China}

\author[Wang]{Shouhong Wang}
\address[SW]{Department of Mathematics,
Indiana University, Bloomington, IN 47405}
\email{showang@indiana.edu}

\thanks{The work was supported in part by the
Office of Naval Research and by the National Science Foundation.}


\begin{abstract}
In this article, we derive a rigorous characterization of the boundary-layer and interior separations in the Taylor-Couette-Poiseuille (TCP) flow. The results obtained  provide a rigorous characterization on how, when and where the propagating Taylor vortices (PTV) are generated. In particular, contrary to what is commonly believed, we show that the 
PTV do not appear after the first dynamical bifurcation, and they appear only when the Taylor number is further increased to cross another critical value so that a structural bifurcation occurs. This structural bifurcation corresponds to the boundary-layer and interior  separations of the flow structure in the physical space.  
\end{abstract}
\maketitle

\section{Introduction}
\label{sc1}
Consider a viscous fluid between two coaxial rotating cylinders. The 
base (Couette) flow becomes unstable as soon as the rotation 
speed of the inner cylinder exceeds a critical value. This instability 
gives rise to a stationary axisymmetric counter-rotating vortices 
that fill the whole annular region. The associated flow is referred to as Taylor--Couette (TC) flow. When a through flow driven by a pressure gradient along the rotating axis is added, the resulting system can exhibit both convective and absolute instabilities.
The base flow consists of a superposition of circular Couette flow 
and annular Poiseuille flow, called Couette-Poiseuille (CP) flow. 
The axial through--flow suppresses the basic stationary instability, 
and as the axial pressure gradient increases while the rotation speed 
of the inner cylinder is held fixed, the first bifurcation gives 
rise to a traveling train of axisymmetric Taylor vortices,
commonly referred to as propagating Taylor vortices (PTV). 
Henceforth, the term Taylor--Couette--Poiseuille (TCP) flow is 
used to refer to all hydrodynamic phenomena
pertaining to the open system described above; see among others 
Raguin and Georgiadis \cite{rg} and the references therein.

In this article we rigorously characterize the stability and  transitions of the CP and TCP flows in both the physical and phase spaces. The main focus is on  a rigorous characterization on how, when and where the PTV are generated. 

We first examine the existence, explicit formula, and basin of attractions of  the secondary flows. The existence of a bifurcation to a steady state solution is classical, and can be proved by the classical Krasnoselskii theorem. The new  ingredients here are the4 explicit formula and the basin of attraction  of the bifurcated solutions. These  enable us to determine, in the next two sections,  the asymptotic structure in the physical space of the solutions of the problem, leading to justification of the flow structure of the  TCP flow.  

The main result of this article is on the rigorous characterization of the boundary-layer and interior  separations, associated with the TCP flows. 
In particular, contrary to what is commonly believed, we show that the 
PTV do not appear after the first dynamical bifurcation, and they appear only when the Taylor number is further increased to cross another critical value so that a structural bifurcation occurs. This structural bifurcation corresponds to the boundary layer and interior  separations of the flow structure in the physical space.  This study gives another example linking the dynamics of fluid flows to the structure and its transitions in the physical spaces.

The analysis is based on a geometric  theory  of two-dimensional (2D) incompressible flows,  and a bifurcation theory for nonlinear partial differential equations and both developed recently by the authors; see  respectively \cite{b-book} and \cite{amsbook} and the references therein.

The geometric theory of 2D incompressible flows was initiated by the authors to study the structure and its stability and transitions of 2-D incompressible fluid flows in the physical spaces. This program of study consists of research in two directions: 
1) the study of the structure and its transitions/evolutions of divergence-free vector fields, and 
2) the study of the structure and its transitions of the flow fields of fluid flows governed 
by the Navier-Stokes equations or the Euler equations. The study in the first direction  is 
more kinematic in nature, and the results and methods developed can naturally be applied to other problems of mathematical physics involving divergence-free vector fields. In fluid dynamics context, the study in  the second direction involves specific connections between the solutions of the Navier-Stokes or the Euler equations and flow structure in the physical space. In other words, this area of research links the kinematics to the dynamics of fluid flows. This is unquestionably an important and difficult problem.
Progresses have been made in several directions. First, a new rigorous characterization of boundary-layer separations for 2D viscous incompressible flows is developed recently by the authors, in collaboration in part with Michael Ghil; see \cite{amsbook} and the references therein. 
Another example in this area is the structure (e.g. rolls) in the physical space in the Rayleigh-B\'enard convection, using the structural stability theorem developed in Area 1) together with application of the aforementioned bifurcation theory; 
see \cite{benard, amsbook}. We would like to mention that this article gives another example in this program of research.

The dynamic bifurcation theory is centered at a new notion of bifurcation, called attractor bifurcation for dynamical systems, both finite dimensional and infinite dimensional, together with new  new strategies for the Lyapunov-Schmidt reduction and the center manifold reduction procedures. The bifurcation theory has been applied to various problems from science and engineering, including, in particular, the Kuramoto-Sivashinshy equation, the Cahn-Hillard equation, the Ginzburg-Landau equation, Reaction-Diffusion equations in Biology and Chemistry, the B\'enard convection problem and the Taylor problem in classical fluid dynamics, and the some geophysical fluid dynamics problems. We mention the interested readers to a recent monograph by the authors \cite{b-book}  and the references therein.

The paper is organized as follows. Section 2 introduces the physical problem, TCP problem. Section 3 proves an abstract dynamical bifurcation theorem, which will be used in Section 4 to derive an explicit form and the basin of attractions of the bifurcated solutions. Sections 5 and 6 characterize the boundary-layer and interior separations associated with the TCP flow.

\section{Couette-Poiseuille flow}
\label{sc2.1}
\subsection{Governing equations}
We consider the viscous flow between two rotating coaxial cylinders
with an axial pressure gradient. The instability of
this flow was first discussed by S.\ Goldstein in 1933 \cite{goldstein}, and
a further investigation was made by S.\ Chandrasekhar~\cite{chandrasekhar}.

Let $r_1$ and $r_2$ $(r_1<r_2)$ be the radii of two coaxial
cylinders, $\Omega_1$ and $\Omega_2$ the angular velocities with
which the  inner and the outer cylinders rotate respectively, and
\begin{equation}
\label{eq2.1}
\mu = \Omega_2/\Omega_1, \qquad \eta = r_1/r_2.
\end{equation}

The hydrodynamical equations governing an incompressible viscous
fluid between two coaxial cylinders are the Navier--Stokes equations
in the cylinder polar coordinates $(z,r,\theta)$, given by 
\begin{equation}
\label{eq2.2}
\left\{ \begin{aligned}
&\frac{\partial u_z}{\partial t} + (u\cdot \nabla ) u_z = \nu\Delta
u_z - \frac{1}{\rho} \, \frac{\partial p}{\partial z}\, , \\
&\frac{\partial u_r}{\partial t} + (u\cdot \nabla ) u_r -
\frac{u^2_\theta}{r} = \nu\left(\Delta u_r - \frac{2}{r^2}\,
\frac{\partial u_\theta}{\partial \theta} - \frac{u_r}{r^2}\right) -
\frac{1}{\rho} \, \frac{\partial p}{\partial  r}\, \\
&\frac{\partial u_\theta}{\partial t} + (u\cdot \nabla ) u_\theta +
\frac{u_ru_\theta}{r} = \nu \left(\Delta u_\theta + \frac{2}{r^2} \,
\frac{\partial u_r}{\partial \theta} - \frac{u_\theta}{r^2}\right) -
\frac{1}{r\rho}\, \frac{\partial p}{\partial \theta}\, , \\
&\frac{\partial (ru_z)}{\partial z} + \frac{\partial (ru_r)}{\partial
r} + \frac{\partial u_\theta}{\partial \theta} = 0,
\end{aligned} \right.
\end{equation}
where $\nu$ is the kinematic viscosity, $\rho$ the density,
$u=(u_r,u_\theta,u_z)$ the velocity field, $p$ the pressure function,
and
\begin{eqnarray*}
u \cdot \nabla  &=& u_z \frac{\partial }{\partial z} + u_r
\frac{\partial }{\partial r} + \frac{u_\theta}{r}\, \frac{\partial
}{\partial \theta}\, , \\
\Delta &=& \frac{\partial ^2}{\partial z^2} + \frac{\partial
^2}{\partial r^2} + \frac{1}{r} \, \frac{\partial }{\partial r} +
\frac{1}{r^2}\, \frac{\partial ^2}{\partial \theta^2}\, .
\end{eqnarray*}

In this article, we consider the case where 
a constant pressure gradient
$\frac{\partial p}{\partial z} = p_0$ is applied in the
$z$-direction, the cylinders rotate, and there are no radial motions. 
Under these conditions, 
 (\ref{eq2.2}) admits a steady state
solution:
\begin{equation}
\label{eq2.3}
\mathcal U^{cp}= (u_z,u_r,u_\theta) = (W,0,V), \qquad p = p(r,z),
\end{equation}
where 
\begin{align}
& \label{eq2.4}
\frac{1}{\rho}\, \frac{dp}{dr} = V^2/r,\\
& 
\label{eq2.5}
\nu\,  \frac{d}{dr} \left( \frac{d}{dr} + \frac{1}{r}\right) V =0, \\
& 
\label{eq2.6}
\nu \left( \frac{d}{dr} + \frac{1}{r}\right) \frac{d}{dr}W =
\frac{p_0}{p}, 
\end{align}
supplemented with the following boundary conditions
\begin{equation}
\label{eq2.7}
\left\{ 
\begin{aligned}
& (u_r, u_z) = \left(0, - \frac{p_0}{4\rho\nu}\, W_0\right)   
    &&  \text{at  } r=r_1,r_2, \\
&u_\theta = r_i\Omega_i &&  \text{at  } r=r_i \text{ (i=1, 2).}
\end{aligned} \right.
\end{equation}
Here $W_0 \ge 0$ is a constant. 

We derive then from (\ref{eq2.4})---(\ref{eq2.7})  that
\begin{align}
& \label{eq2.8}
V(r) = \frac{\Omega_1}{1-\eta^2}\left(r^2_1(1-\mu) \frac{1}{r} -
(\eta^2-\mu)r\right), \\
&
\label{eq2.9}
W(r) = -\frac{p_0}{4\rho\nu}\left(r^2_1 - r^2 +
\frac{2r_1d+d^2}{\ln(1+d/r_1)}\, \ln \left(\frac{r}{r_1}\right) +
W_0\right),\\
&
\label{eq2.10}
p(r,z) = p_0z + \rho \int \frac{1}{r}\, V^2(r)dr,
\end{align}
where $d$ $(=r_2-r_1)$ is the gap width, $\mu$ and $\eta$ are as in
(\ref{eq2.1}).

Hereafter, we always assume that
\begin{equation}
\label{eq2.11}
\eta^2>\mu,
\end{equation}
which is a necessary condition for the instability to occur.

Classically, the flow described by $\mathcal U^c=(0,0,V(r))$ is called the
Couette flow, and the flow by $\mathcal U^p=(W(r),0,0)$ is called the Poiseuille
flow. Therefore the flow described by (\ref{eq2.3}) is a
super-position of the Couette flow and the Poiseuille flow, and is 
usually called the Couette-Poiseuille (CP) flow. 

The main objective 
of this article is to study the stability and transitions of the CP flow 
in both the physical and the phase spaces.

For this purpose, we consider the perturbed state
\begin{equation}
\label{eq2.12}
W+u_z,\,\, u_r,\,\, V+u_\theta,\qquad p+p_0z+\rho\int
\frac{1}{r}\, V^2dr.
\end{equation}

Assume that the perturbation is axisymmetric and independent
of~$\theta$, we derive from (\ref{eq2.2}) that
\begin{equation}
\label{eq2.13}
\left\{ \begin{aligned}
&\frac{\partial u_z}{\partial t} + (\widetilde u \cdot \nabla )u_z =
\nu\Delta u_z - W\, \frac{\partial u_z}{\partial z} - u_r\,
\frac{dW}{dr} - \frac{1}{\rho}\, \frac{\partial p}{\partial z}\, , \\
&\frac{\partial u_r}{\partial t} + (\widetilde u \cdot \nabla ) u_r -
\frac{u^2_\theta}{r} = \nu \left( \Delta u_r - \frac{u_r}{r^2}\right)
- \frac{1}{\rho}\, \frac{\partial p}{\partial r}\\
&  \qquad  \qquad \qquad \qquad  - W\, \frac{\partial
 u_r}{\partial z} + \frac{2 V}{r}\, r_\theta, \\
&\frac{\partial u_\theta}{\partial t} + (\widetilde u \cdot \nabla )
u_\theta + \frac{u_\theta u_r}{r} = \nu\left( \Delta u_\theta -
\frac{u_\theta}{r^2}\right) \\
& \qquad \qquad \qquad \qquad - \left(V' + \frac{1}{r}\, V\right) u_r -
W\, \frac{\partial u_\theta}{\partial z}\, , \\
&\frac{\partial (ru_z)}{\partial z} + \frac{\partial (ru_r)}{\partial
r} = 0,
\end{aligned} \right.
\end{equation}
where $\widetilde u = (u_z,u_r)$, and 
\begin{eqnarray*}
&&\Delta = \frac{\partial ^2}{\partial r^2} + \frac{1}{r}\,
\frac{\partial }{\partial r} + \frac{\partial ^2}{\partial z^2}\, ,
\\
&&(\widetilde u \cdot \nabla ) = u_z\, \frac{\partial }{\partial z} +
u_r\, \frac{\partial }{\partial r}\, .
\end{eqnarray*}

The spatial domain for (\ref{eq2.13}) is $M = (0,L) \times(r_1,r_2)$, 
where $L$ is the height of the cylinders. 
The physically sound boundary conditions are as follows
\begin{equation}
\label{eq2.14}
\left\{ \begin{aligned}
&u = (u_z,u_r,u_\theta) = 0 &&  \text{at $r=r_1$, $r_2$,} \\
&u_z =0, \,\,\frac{\partial u_r}{\partial z} = \frac{\partial
u_\theta}{\partial z} =0 \qquad &&\text{at $z=0,L$.}
\end{aligned}  \right.
\end{equation}

\subsection{Narrow-gap approximation}
\label{sc2.3}
Consider the narrow-gap approximation with $\mu \ge 0$. Namely, we 
assume that the gap $d=r_2-r_1$ is small in comparison to the mean
radius:
\begin{equation}
\label{eq2.15}
d = r_2-r_1 \ll \frac{1}{2} (r_2+r_1).
\end{equation}
Under this assumption,  we can neglect the terms having
the factors $r^{-n}$ $(n\ge 1)$ in the equations. Let
\begin{equation}
\label{eq2.16}
\alpha = \frac{\eta^2 -\mu}{1-\eta^2}.
\end{equation}
By (\ref{eq2.11}), $\alpha>0$. We derive then from 
(\ref{eq2.8}) and (\ref{eq2.9})
that 
\begin{align}
& V' + \frac{1}{r}\, V = -2 \alpha\Omega_1, \\
& \frac{2V}{r} = 2\Omega_1 \left( 1- \frac{1-\mu}{1-\eta^2} \,
\frac{r^2-r^2_1}{r^2}\right) \simeq 2\Omega_1 \left( 1- \frac{(1-\mu)(r-r_1)}{d}\right), \\
& W \simeq - \frac{p_0}{4\rho\nu}\left[ (r-r_1)(r_2-r) +W_0\right].
\end{align}

Replacing $u_\theta$ by $\sqrt{\alpha}\, u_\theta$ in (\ref{eq2.13}),
we obtain the following  approximation equations describing the flow
between two cylinders with a narrow gap: 
\begin{equation}
\label{eq2.17}
\left\{ \begin{aligned}
\frac{\partial u_z}{\partial t} + (\widetilde u \cdot \nabla ) u_z =
&\nu \Delta u_z + \frac{p_0}{4\rho \nu}(d-2 \overline{r})u_r \\
& + \frac{p_0}{4\rho\nu}(W_0 + \overline{r}(d-\overline{r}))
\frac{\partial u_z}{\partial z} - \frac{1}{\rho}\, \frac{\partial
p}{\partial z}\, , \\
\frac{\partial  u_r}{\partial t} + (\widetilde u\cdot \nabla ) u_r =
&\nu \Delta u_r + 2 \sqrt{\alpha}\, \Omega_1 \left(1-
\frac{(1-\mu)\overline{r}}{d}\right) u_\theta \\
&+ \frac{p_0}{4\rho\nu}(W_0 + \overline{r}(d-\overline{r}))
\frac{\partial u_r}{\partial z} - \frac{1}{p}\, \frac{\partial
p}{\partial r}\, , \\
\frac{\partial u_\theta}{\partial t} + (\widetilde u \cdot \nabla )
u_\theta = &\nu \Delta u_\theta + 2 \sqrt{\alpha}\, \Omega_1 u_r +
\frac{p_0}{4\rho\nu}(W_0 +
\overline{r}(d-\overline{r}))\frac{\partial u_\theta}{\partial z}\, ,
\\
\divv \, \widetilde u =0,\qquad\quad\,\,\, &
\end{aligned} \right.
\end{equation}
where $\overline{r} = r-r_1$, $\Delta = \partial ^2/\partial  z^2 +
\partial ^2/\partial r^2$.

We need to consider the nondimensionalized form of (\ref{eq2.17}). To
this end, let
\begin{eqnarray*}
x&=& x'd \qquad\quad (x=(z,r,r\theta)), \\
t &=& t'd^2/\nu, \\
u &=& u'\nu/d \qquad (u = (u_z,u_r,u_\theta)), \\
p &=& p' \rho\nu^2/d^2, \\
W_0&=& W_0' d^2.
\end{eqnarray*}
Omitting the primes, the equations (\ref{eq2.17}) can be rewritten as
\begin{equation}
\label{eq2.18}
\left\{ \begin{aligned}
\frac{\partial u_z}{\partial t} = \,&\Delta u_z +
\gamma(1-2\overline{r})u_r \\ 
& + \gamma(W_0 +
\overline{r}(1-\overline{r})) 
 \frac{\partial u_z}{\partial z} - \frac{\partial
p}{\partial z} - (\widetilde u\cdot \nabla )u_z, \\
\frac{\partial u_r}{\partial t} = &\Delta u_r +
\lambda(1-(1-\mu)\overline{r})u_\theta\\
 & - \gamma(W_0 +
\overline{r}(1-\overline{r})) \frac{\partial u_r}{\partial z} -
\frac{\partial p}{\partial r} - (\widetilde u\cdot \nabla )u_r, \\
\frac{\partial u_\theta}{\partial t} = &\Delta u_\theta + \lambda u_r
+ \gamma(W_0 + \overline{r} (1-\overline{r})) \frac{\partial
u_\theta}{\partial z} - (\widetilde u \cdot \nabla )u_\theta, \\
\divv \, \widetilde u =\, &0,
\end{aligned} \right.
\end{equation}
where $\lambda = \sqrt{T}$. Here the Taylor number $T$ and the 
nondimensional parameter  $\gamma$ are given by 
\begin{equation}
\label{eq2.19}
T = \frac{4\alpha\Omega^2_1 d^4}{\nu^2}\, ,\qquad \gamma =
\frac{p_0d^3}{4\rho\nu^2}\,. 
\end{equation}
We remark here that the nondimensional parameter $\gamma$, proportional to the Reynolds number $Re$ squared times the small gap $d$, is a small parameter.

When the gap $d=r_2-r_1$ is small in comparison 
with the mean radius, we
need not to distinguish the two equations (\ref{eq2.13}) and
(\ref{eq2.18}) to discuss their dynamic bifurcation. Therefore, we
always consider the problem (\ref{eq2.18}) with (\ref{eq2.14}) 
instead of
(\ref{eq2.13}) with (\ref{eq2.14}). The equations (\ref{eq2.18})
are supplemented with the following initial condition
\begin{equation}
\label{eq2.20}
u=\varphi  \quad \text{at}\quad t=0.
\end{equation}

\section{Dynamic Bifurcation for Perturbed Systems}
\label{sc4}

\subsection{Preliminaries}
\label{sc4.1}
Let $H_1$ and $H$ be two Hilbert spaces, and $H_1\to H$ be a dense
and compact inclusion. We consider the following nonlinear evolution
equation:
\begin{equation}
\label{eq4.1}
\left\{ \begin{aligned}
&\frac{du}{dt} = Au + \lambda Bu + {\cL}^\varepsilon_\lambda u +
G(u,\lambda), \\
&u(0) = \varphi,
\end{aligned} \right.
\end{equation}
where $u:[0,\infty) \to H$ is the unknown function, $\lambda\in \R$
and $\varepsilon\in \R^m$ $(m\ge 1)$ are the system parameters, and
$A,B:H_1\to H$ are linear operators, $ \cL^\varepsilon_\lambda:
H_1\to H$ are parameterized linear operators continuously depending
on $\lambda\in R$, $\varepsilon\in \R^m$, which satisfy
\begin{align}
&
\label{eq4.2}
\left\{ \begin{aligned}
&A:H_1\to H  &&   \text{a sectorial operator,} \\
&A^\alpha:H_\alpha\to H  &&  \text{the fractional power operators
for $\alpha\in \R$,} \\
&H_\alpha =D(A^\alpha) &&  \text{the domain of $A^\alpha$ with
$H_0 =H$;}
\end{aligned} \right.
\\
&
\label{eq4.3}
\left\{ \begin{aligned}
&\cL^\varepsilon_\lambda,B:H_\theta\to H &&  \text{bounded for some
$\theta<1$, } \\
&\|\cL^\varepsilon_\lambda\|\to 0  &&  \text{if $\varepsilon\to0$,
$\forall$ $\lambda\in \R$.}
\end{aligned}\right.
\end{align}

We know that $H_\alpha \subset H_\beta$ are compact inclusions for
all $\alpha>\beta$, and the operators $A+\lambda
B+\cL^\varepsilon_\lambda:H_1\to H$ are sectorial operators.
Furthermore, we assume that the nonlinear terms $G(\cdot,\lambda):
H_\alpha \to H$ for some $\alpha<1$ are a family of parameterized
$C^r$ bounded operators $(r\ge 1)$, depending continuously on the
parameter $\lambda\in \R$, such that
\begin{equation}
\label{eq4.4}
G(u,\lambda) =o\left( \|u\|_{H_\alpha}\right), \quad \forall\,\,
\lambda\in \R.
\end{equation}

Let $\{S(t)\}_{t\ge 0}$ be an operator semigroup generated by
(\ref{eq4.1}), which enjoys the semigroup properties.
Then the solution of (\ref{eq4.1}) can be expressed as
$u(t,\varphi) = S(t)\varphi$ for any  $t\ge 0.$

\begin{definition}
\label{df4.1}
A set $\Sigma \subset H$ is called an invariant set of (\ref{eq4.1})
if $S(t) \Sigma = \Sigma$ for any $t\ge 0$. An invariant set
$\Sigma\subset H$ of (\ref{eq4.1}) is called an attractor if $\Sigma$
is compact, and there exists a neighborhood $U\subset H$ of $\Sigma$
such that for any $\varphi \in U$ we have
\[
\lim_{t\to \infty} \dist (S(t)\varphi,\Sigma) =0 \quad \text{in $H$
norm,}
\]
and we say that $\Sigma$ attracts $U$. Furthermore, if $U=H$, then
$\Sigma$ is called a global attractor of (\ref{eq4.1}).
\end{definition}

A number $\beta=\beta_1 + i\beta\in \C$ is called an eigenvalue of a
linear operator $L:H_1\to H$ if there exist $x,y\in H_1$ with $x\not=
0$ such that
\[
Lz=\beta z, \qquad (z = x+iy).
\]
The space
\[
E_\beta = \bigcup_{n\in \N} \left\{ x,y \in H_1 \mid (L-\beta)^nz =0,
\,\,z =x +iy\right\}
\]
is called the eigenspace of $L$, and $x,y \in E_\beta$ are called
eigenvectors of $L$. 

\begin{definition}
\label{df4.2}
A linear mapping $L^* :H_1 \to H$ is called the conjugate operator of
$L: H_1\to H$, if
\[
\langle Lx,y\rangle_H = \langle x,L^*y\rangle_H, \qquad \forall\,\,
x,y \in H_1.
\]
A linear operator $L: H_1\to H$ is called symmetric if $L=L^*$.
\end{definition}

\subsection{Eigenvalue problem}
\label{sc4.2}
Here we consider the eigenvalue problem for the perturbed linear
operators
\begin{align*}
&L^\varepsilon_\lambda = L_\lambda + \cL^\varepsilon_\lambda, \\
&L_\lambda = A + \lambda B.
\end{align*} 
Let the eigenvalues (counting multiplicities) of $L_\lambda$ and
$L^\varepsilon_\lambda$ be given respectively by
\begin{eqnarray*}
&&\left\{ \beta_k(\lambda) \mid k=1,2,\dots\,\right\} \subset \C, \\
&&\left\{\beta^\varepsilon_k(\lambda) \mid k=1,2,\dots\, \right\}
\subset \C.
\end{eqnarray*}

Suppose that the first eigenvalue $\beta_1(\lambda)$ of $L_\lambda$
is simple near $\lambda=\lambda_0$, and the eigenvalues
$\beta_k(\lambda)$ $(k=1,2,\dots\,)$ satisfy the principle of exchange of stabilities (PES):
\begin{align}
&
\label{eq4.5}
\beta_1(\lambda) = \left\{ 
\begin{aligned}
<0  && \text{if }  \lambda<\lambda_0, \\
=0  && \text{if }  \lambda=\lambda_0, \\
>0  && \text{if }  \lambda>\lambda_0,
\end{aligned} \right.
\\
&
\label{eq4.6}
Re\beta_j(\lambda_0)<0  \qquad \forall\,\, j \ge 2.
\end{align}
Let $e_\lambda,e^*_\lambda \in H_1$ be the eigenvectors of
$L_\lambda$ and $L^*_\lambda$ corresponding to an eigenvalue
$\beta(\lambda)$ respectively, i.e.,
\begin{equation}
\label{eq4.7}
\left\{ \begin{aligned}
&Ae_\lambda + \lambda Be_\lambda = \beta(\lambda) e_\lambda, \\
&A^*e^*_\lambda + \lambda B^* e^*_\lambda = \beta(\lambda)
e^*_\lambda.
\end{aligned} \right.
\end{equation}
We also assume that
\begin{equation}
\label{eq4.8}
\langle e_\lambda,e^*_\lambda\rangle_H =1.
\end{equation}

The following theorem is important for our discussion later, which
provides a differential formula for simple eigenvalues of linear
operators.

\begin{theorem}
\label{th4.3}
Let the eigenvalue $\beta(\lambda)$ in (\ref{eq4.7}) be simple. Then
$\beta(\lambda)$ is differentiable at $\lambda$, and with the
assumption (\ref{eq4.8}) we have
\begin{equation}
\label{eq4.9}
\beta'(\lambda) = \langle Be_\lambda,e^*_\lambda\rangle_H.
\end{equation}
\end{theorem}

\begin{proof}
By the genericity of simple eigenvalues of linear operators (see Kato
\cite{kato} and Ma and Wang \cite{b-book}), there is a number $\delta_0>0$
such that as $|\delta|<\delta_0$, $\beta(\lambda+\delta)$ is also a
simple eigenvalue of $L_{\lambda+\delta}$. Let
\begin{equation}
\label{eq4.10}
\left\{ \begin{aligned}
&Au+(\lambda+\delta)Bu = \beta(\lambda+\delta)u, \\
&u=e_\lambda + v_\delta, \\
&\|v_\delta\|_H \longrightarrow 0 \quad \text{ if }\delta\to 0.
\end{aligned} \right.
\end{equation}

We infer from (\ref{eq4.7}), (\ref{eq4.8}) and (\ref{eq4.10}) that
\begin{eqnarray*}
&&\langle (A+\lambda B)e_\lambda,e^*_\lambda\rangle_H +
\langle(A+\lambda B)v_\delta,e^*_\lambda\rangle_H + \delta \langle B
u,e^*_\lambda\rangle_H \\
&=& \beta(\lambda) + \beta(\lambda) \langle v_\delta,e^*_\lambda\rangle_H
+ \delta\langle Bu,e^*_\lambda\rangle_H \\
&=& \beta(\lambda+\delta) + \beta(\lambda+\delta)\langle
v_\delta,e^*_\lambda\rangle_H.
\end{eqnarray*}
Thus the last equality implies  that
\[
\frac{\beta(\lambda+\delta) -\beta(\lambda)}{\delta} \, (1+\langle
v_\delta,e^*_\lambda\rangle_H) = \langle B
(e_\lambda+v_\delta),e^*_\lambda\rangle_H. \]
Then the theorem follows from (\ref{eq4.10}).
\end{proof}

\begin{remark}
\label{rm4.4}
{\rm
For general linear operators $L_\lambda:H_1\to H$, if $L_\lambda$ are
differentiable on $\lambda$ with $\cL_\lambda = \frac{d}{d\lambda}\,
L_\lambda:H_1\to H$ bounded, then Theorem~\ref{th4.3} holds true as well,
and
\begin{equation}
\label{eq4.11}
\beta'(\lambda) = \langle \cL_\lambda e_\lambda,e^*_\lambda\rangle_H.
\end{equation}
\qed}
\end{remark}

By Theorem~\ref{th4.3} or (\ref{eq4.11}), for the perturbed linear
operators $L^\varepsilon_\lambda$ we immediately obtain the following
corollary.

\begin{corollary}
\label{co4.5}
Assume the PES (\ref{eq4.5}) and (\ref{eq4.6}), and 
$\frac{d}{d\lambda}\, \cL^\varepsilon_\lambda$ exists. 
Then for each $|\varepsilon|<\delta$  for some  $\delta>0$, there
exists $\lambda^\varepsilon_0$, with $\lim_{\varepsilon\to0}\lambda^\varepsilon_0 =
\lambda_0$, such that the eigenvalues
$\beta^\varepsilon_k(\lambda)$ of $L^\varepsilon_\lambda$ at
$\lambda^\varepsilon_0$ satisfy the following PES:
\begin{align}
&
\label{eq4.12}
\beta^\varepsilon_1(\lambda)  \left\{
\begin{aligned}
<0 &\quad \text{if $\lambda<\lambda^\varepsilon_0$,} \\
=0 &\quad \text{if $\lambda = \lambda^\varepsilon_0$,} \\
>0 &\quad \text{if $\lambda>\lambda^\varepsilon_0$,}
\end{aligned} \right.
\\
&
\label{eq4.13}
Re\beta^\varepsilon_j(\lambda^\varepsilon_0)<0, \qquad \forall\,\,
2\le j.
\end{align}
Moreover, $\beta^\varepsilon_1(\lambda)$ has the following expansion at
$\lambda= \lambda^\varepsilon_0$:
\begin{equation}
\label{eq4.14}
\left\{ \begin{aligned}
& \beta^\varepsilon_1(\lambda) =
\alpha_\varepsilon(\lambda-\lambda^\varepsilon_0) +
o(|\lambda-\lambda^\varepsilon_0|), \\
& \alpha_\varepsilon = \langle (B + \textstyle{ \frac{d}{d\lambda}}\,
\cL^\varepsilon_{\lambda^\varepsilon_0})e_\varepsilon,
e^*_\varepsilon\rangle_H,
\end{aligned} \right.
\end{equation}
where $e_\varepsilon$ and $e^*_\varepsilon$ satisfy
\begin{equation}
\label{eq4.15}
L^\varepsilon_{\lambda^\varepsilon_0}e_\varepsilon =0, \qquad
L^{\varepsilon^*}_{\lambda^\varepsilon_0}e^*_\varepsilon =0.
\end{equation}
\end{corollary}

\subsection{Bifurcation for perturbed equations}
\label{sc4.3}
We study now the dynamic bifurcation for the nonlinear
evolution equation (\ref{eq4.1}), which is  a perturbation of the
following equation
\begin{equation}
\label{eq4.16}
\frac{du}{dt} = Au +\lambda Bu+G(u,\lambda).
\end{equation}

Under the conditions (\ref{eq4.7}) and (\ref{eq4.8}), let $e_0 \in
H_1$ be  an  eigenvector of $L_\lambda$ at $\lambda=\lambda_0$:
\begin{equation}
\label{eq4.17}
Ae_0 + \lambda_0 Be_0 =0, \qquad \|e_0\|=1.
\end{equation}
We assume that
\begin{equation}
\label{eq4.18}
\left\{ \begin{aligned}
&L_\lambda = A+\lambda B && \text{are symmetric,} \\
&G(u,\lambda) &&  \text{are bilinear,} \\
&\langle G(u,v),v\rangle_H =0 \qquad && \forall\,\, u,v \in H_\alpha.
\end{aligned} \right.
\end{equation}
Here if $G$ is bilinear, one can write $G$ as $G(\cdot,\cdot)$, which
is linear for each argument.

By Theorem~6.15 and Remark~6.8 in \cite{b-book}, together with
Corollary~\ref{co4.5} above, we obtain immediately the following theorem.

\begin{theorem}
\label{th4.6} Assume that (\ref{eq4.2})---(\ref{eq4.6}) and (\ref{eq4.18}) hold true, and
$u=0$ is a globally asymptotically stable equilibrium point of
(\ref{eq4.16}) at $\lambda=\lambda_0$. Then there are constants $\delta_1>0$
and $\delta_2>0$ such that if $|\varepsilon|<\delta_1$ and
$0<\lambda-\lambda^\varepsilon_0<\delta_2$, where
$\lambda^\varepsilon_0$ is given in (\ref{eq4.12}) and (\ref{eq4.13}), then the
following assertions hold true.
\begin{enumerate}
\item There exists an attractor $\Sigma^\varepsilon_\lambda = \{u^\lambda_1,u^\lambda_2\}\subset H$
of (\ref{eq4.1}), where $u^\lambda_1$  and $u^\lambda_2$  are steady states given by 
\begin{equation}
\label{eq4.19}
\left\{ \begin{aligned}
&u^\lambda_1 = \sigma_1(\lambda,\varepsilon)e_0 +
v_1(\lambda,\varepsilon), \\
&u^\lambda_2 = -\sigma_2(\lambda,\varepsilon)e_0
+v_2(\lambda,\varepsilon), \\
&v_i(\lambda,\varepsilon) = o(|\sigma_i(\lambda,\varepsilon)|) \in
H_1, \quad i =1,2.
\end{aligned} \right.
\end{equation}
Here $\sigma_i(\lambda,\varepsilon) \in \R$ can be expressed as
\begin{equation}
\label{eq4.20}
\left\{ \begin{aligned}
&\sigma_{1,2} = \frac{\sqrt{b^2(\varepsilon) +
4\beta^\varepsilon_1(\lambda)} \mp |b(\varepsilon)|}{2C} +
o(|b|,|\beta^\varepsilon_1|), \\
&b(\varepsilon) = \langle G(e_\varepsilon,
\lambda^\varepsilon_0),e^*_\varepsilon\rangle_H,
\end{aligned} \right.
\end{equation}
where $e_\varepsilon,e^*_\varepsilon$ are as in
(\ref{eq4.15}), $\beta^\varepsilon_1(\lambda)$ as in (\ref{eq4.14}),
and $C>0$ is a constant.

\item $\Sigma^\varepsilon_\lambda $  attracts every bounded  open set in 
$H \setminus \Gamma$, where
$\Gamma$ is the stable manifold of $u=0$ with co-dimension one. Namely, 
 $H$ can be decomposed into two open sets $U^\lambda_1$ and
$U^\lambda_2$:
\[
H = \overline{U}^\lambda_1 + \overline{U}^\lambda_2, \qquad
U^\lambda_1 \cap U^\lambda_2 = \emptyset, \qquad 0\in \Gamma=
\partial U^\lambda_1 \cap \partial U^\lambda_2, 
\]
with $u^\lambda_i \in U^\lambda_i$ $(i=1,2)$, such that
\[
\lim_{t\to \infty} \|u(t,\varphi) - u^\lambda_i\|_H =0, \quad
\text{if $\varphi \in U^\lambda_i$ $(i=1,2)$,}
\]
where $u(t,\varphi)$ is the solution of (\ref{eq4.1}).
\end{enumerate}

\end{theorem}

A few remarks are now in order.

\begin{remark}
\label{rm4.7}
{\rm 
It is clear that the eigenvectors in (\ref{eq4.15}) satisfy that
$e_\varepsilon \to e_0$, $e^*_\varepsilon \to e_0$ when
$\varepsilon\to 0$, where $e_0$ is as in (\ref{eq4.17}). Therefore, by
(\ref{eq4.18}) we have $b(\varepsilon) \to0$ as $\varepsilon\to0$.
}
\end{remark}

\begin{figure}
 \centering \includegraphics[height=0.5\hsize]{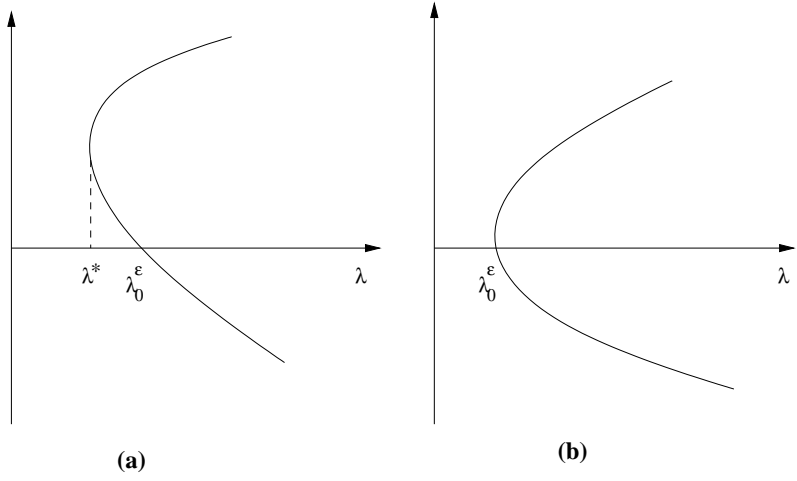}
\caption{(a) Saddle-node bifurcation and hysteresis, and (b) supercritical attractor bifurcation.}
\label{fg4.1}
\end{figure}

\begin{remark}
\label{rm4.8}
{\rm
For a given $\varepsilon>0$, if $b(\varepsilon) \not= 0$, then the
equation (\ref{eq4.1}) has a saddle--node bifurcation at
$\lambda^*<\lambda^\varepsilon_0$  as shown in Figure~\ref{fg4.1}(a); 
if $b(\varepsilon) =0$, then (\ref{eq4.1}) has an (supercritical) attractor
bifurcation  as shown in Figure~\ref{fg4.1}(b).
}
\end{remark}

To apply the above dynamic bifurcation theorem it is crucial to
verify the global asymptotic stability for (\ref{eq4.16}) at
$\lambda=\lambda_0$. The following theorem plays an important role to
ensure this condition, which was proved in \cite{benard, b-book}.

\begin{theorem}
\label{th4.9}
Under the conditions (\ref{eq4.5}), (\ref{eq4.6}) and (\ref{eq4.18}),
if $G(e_0,\lambda_0) \not= 0$, then $u=0$ is a globally asymptotically
stable equilibrium point of (\ref{eq4.16}) at $\lambda=\lambda_0$.
\end{theorem}

\section{Dynamic bifurcation and stability of the Couette-Poiseuille Flow}
\label{sc5}
\subsection{Functional setting}
We recall here the functional setting and some basic
mathematical properties of equations (\ref{eq2.18}) with boundary
conditions given by (\ref{eq2.14}). The spatial domain is $M =
(r_1,r_1+1)\times (0,L)$, the coordinate system is $(r,z)$, the
velocity field is $u = (u_z,u_r,u_\theta)$ and $\widetilde u =
(u_z,u_r)$. Let
\begin{equation}
\label{eq5.1}
\left\{ \begin{aligned}
&H=\{u = (\widetilde u, u_\theta) \in L^2(M)^3\mid \divv\, \widetilde
u =0,\,\, \widetilde u\cdot n|_{\partial M} =0\}, \\
&V= \{ u\in H^1(M)^3 \cap H\mid u =0 \,\,\text{at $r=r_1,r_2$,
$u_z
=0$ at $z=0,L$}\}, \\
&H_1 = \{ u\in H^2(M)^3 \cap V \mid \textstyle{\frac{\partial
u_r}{\partial z} = \frac{\partial u_\theta}{\partial z} = 0}
\,\,\text{at $z=0,L$}\},
\end{aligned} \right.
\end{equation}
where $H^k(M)$ is the usual Sobolev spaces.

The equations (\ref{eq2.18}) are two--dimensional, therefore the
results concerning the existence and regularity of a solution for
(\ref{eq2.18}) with (\ref{eq2.14}) are classical. For each initial
value $u_0 \in H$, (\ref{eq2.18})  and  (\ref{eq2.20}) with
(\ref{eq2.14}) has a weak solution
\[
u\in L^\infty ([0,T],H) \cap L^2 ([0,T],V), \qquad \forall\,\, T>0.
\]
If $u_0\in V$, (\ref{eq2.18}), (\ref{eq2.20}) with (\ref{eq2.14}) has
a unique solution
\[
u \in C([0,T],V) \cap L^2 ([0,T],H_1),\qquad \forall \,\, T>0.
\]

For every $u_0 \in H$ and $k\ge 1$ there exists a $\tau_0>0$ such
that the solution $u(t,u_0)$ of (\ref{eq2.18}), (\ref{eq2.20}) with
(\ref{eq2.14}) satisfies
\[
u(t,u_0)\in H^k (M)^3, \qquad \forall\,\, t>\tau_0.
\]
Furthermore, for each $k\ge 1$ and $u_0 \in H^k(M)^3$, there exists a
number $C>0$ depending on $k$ and $\varphi$ which bounds the solution
$u(t,u_0)$ of (\ref{eq2.18}) and (\ref{eq2.20}) with (\ref{eq2.14}) in
the $H^k$--norm (see \cite{fmt}):
\[
\|u(t,u_0)\|_{H^k} \le C, \qquad \forall\,\, t\ge 0.
\]

Finally, it is essentially known that for any $\lambda,\gamma$ and
$\mu$ the equations (\ref{eq2.18}) with (\ref{eq2.14}) have a global
attractor; see Temam \cite{temam} and Foias {\it et al.}~\cite{fmt}.

\subsection{Eigenvalue problem of the linearized equations}
\label{sc5.2}
The linearized equations of (\ref{eq2.18}) with (\ref{eq2.14}) are
given by
\begin{equation}
\label{eq5.2}
\left\{ \begin{aligned}
&-\Delta u_z - \gamma(1-2\overline{r})u_r + \gamma(W_0 +
\overline{r}(1-\overline{r})) \frac{\partial u_z}{\partial z} +
\frac{\partial p}{\partial z} =0, \\
&-\Delta u_r + \lambda(1-\mu)\overline{r} u_\theta + \gamma(W_0 +
\overline{r}(1-\overline{r})) \frac{\partial u_r}{\partial z} +
\frac{\partial p}{\partial r} =
\lambda u_\theta, \\
&-\Delta u_\theta - \gamma(W_0 + \overline{r}(1-\overline{r}))
\frac{\partial u_\theta}{\partial z} = \lambda u_r, \\
&\quad \frac{\partial u_z}{\partial z} + \frac{\partial
u_r}{\partial r} =0,
\end{aligned} \right.
\end{equation}
with the boundary conditions
\begin{equation}
\label{eq5.3}
\left\{ \begin{aligned}
&u=0  && \text{at $r=r_1$, $r_1+1$,} \\
&u_z =0, \quad  \frac{\partial u_r}{\partial z} = \frac{\partial
u_\theta}{\partial z} =0 &&  \text{at $z=0,L$,}
\end{aligned} \right.
\end{equation}
where $\overline{r} = r-r_1$, $\lambda = \sqrt T$, $T$ is the Taylor
number, $T$  and $\gamma$ are given by (\ref{eq2.19}), and $\mu$ is defined by 
(\ref{eq2.1}). We denote
\[
\varepsilon = (\gamma,1-\mu) \in (0,\infty) \times (0,\infty). \]

Let $\lambda^\varepsilon_0 >0$ be the first eigenvalue of
(\ref{eq5.2}), (\ref{eq5.3}), and we call
\begin{equation}
\label{eq5.4}
T_c = (\lambda^{\varepsilon}_0)^2
\end{equation}
the critical Taylor number.

When $\varepsilon\to 0$, the system (\ref{eq5.2}) reduces to the
following symmetric linear equations
\begin{equation}
\label{eq5.5}
\left\{ \begin{aligned}
&-\Delta u_z + \frac{\partial p}{\partial z} =0, \\
&-\Delta u_r + \frac{\partial p}{\partial r} = \lambda u_\theta, \\
&-\Delta u_\theta = \lambda u_r, \\
&\quad \frac{\partial u_z}{\partial z} + \frac{\partial
u_r}{\partial r} =0.
\end{aligned} \right.
\end{equation}

It is known that the eigenvalue problem (\ref{eq5.5}) with boundary
conditions (\ref{eq5.3}) has the eigenvectors as follows
\begin{align}
\label{eq5.6}u_0& =(u_z, u_r, u_\theta)\\
& =(\frac{1}{a}\, \sin az R'(r),\frac{1}{a}\, \sin az R'(r),
 \frac{-1}{a^2\lambda_0}\, \cos az \left( \frac{d^2}{dz^2}
- a^2\right)^2R(r)), \nonumber 
\end{align}
where $a$ is the wave length which is given by (see \cite{chandrasekhar})
\[
a = \frac{K \pi}{L}, 
\]
for some $K\ge 1$, with $K$ depending on $L$. When $L$ is large, $K$ is taken to be 
the minimizer of 
\[
\min_k \left| \frac{k\pi}{L} - 3.117\right|.
\]
The number $\lambda_0$ is the first eigenvalue of (\ref{eq5.5}), 
and when $a \simeq
3.117$, $\lambda^2_0 \simeq 1700$.

Consider 
\[
\left\{ \begin{aligned}
&\left( \frac{d^2}{dr^2} - a^2\right)^3R = -a^2\lambda^2_0R, \\
&R=0, \,\, R'=0, \,\,\left( \frac{d^2}{dr^2} - a^2\right)^2R=0
\,\,\text{at $r=r_1$, $r_1+1$. }
\end{aligned} \right.
\]
Its solution can be expressed as (see Chapter II--15 of \cite{chandrasekhar})
\begin{equation}
\label{eq5.7} R(r) = \cos \alpha_0 x-\beta_1 \cosh \alpha_1x+
\cos \alpha_2x + \beta_2 \sinh\alpha_1 x \sin \alpha_2 x,
\end{equation}
where $x = \overline{r} - \frac{1}{2}$, $\overline{r} = r-r_1$,
\begin{equation}
\label{eq5.8}
\left\{ \begin{aligned}
&\beta_1 = 0.06151664, \quad \beta_2 = 0.10388700, \\
&\alpha_0 = 3.973639, \quad \alpha_1 = 5.195214, \quad \alpha_2 =
2.126096.
\end{aligned} \right.
\end{equation}

In addition, the first eigenvalue $\lambda_0$ of (\ref{eq5.5}) with
(\ref{eq5.3}) is simple except for the following values of $L$:
\begin{equation}
\label{eq5.9}
L_k = \frac{k\pi}{\alpha_0}\, , \quad k=1,2,\cdots, \quad a_0 \simeq
4.2.
\end{equation}

\subsection{Secondary flows and their stability}
\label{sc5.3}
We are now in  position to state and prove the main
theorem in this section. We remark here that under conditions
(\ref{eq2.11}) and (\ref{eq2.15}) with scaling $d=1$, the condition
$\mu\to 1$ can be replaced by
\begin{equation}
\label{eq5.10}
(1-\mu)r_1 = 2+\delta
\end{equation}
for some $\delta>0$ and $r_1$ sufficiently large. This condition
implies that 
$$\alpha = (\eta^2-\mu)/(1-\eta^2) \simeq \delta.$$

\begin{theorem}
\label{th5.1}
Assume that (\ref{eq5.10}) holds true, and  $L\not= L_k$ is
sufficiently large, where $L_k$ is given by (\ref{eq5.9}). 
Then there are $R>0$  and
$\delta_1>0$ such that
for any $r_1>R$ and $|\gamma|<\delta_1$, the following assertions hold true,  
provided that  the Taylor number $T=\lambda^2$ satisfies that 
$0<T-T_c<\delta_2$ with $T_c$ as in (\ref{eq5.4})  for some $\delta_2>0$, or equivalently $0<\lambda-\lambda^\varepsilon_0 <\tilde \delta_2$ 
for some $\tilde \delta_2>0$:
 
\begin{enumerate}

\item There exists an attractor 
$\Sigma^\varepsilon_\lambda=\{u^\lambda_1,u^\lambda_2\}\subset H$
of (\ref{eq2.18}) with (\ref{eq2.14}), where $u^\lambda_1$  and $u^\lambda_2$ are steady state solutions given by 
\begin{equation}
\label{eq5.11}
\left\{ \begin{aligned}
&u^\lambda_1 = \sigma_1(\lambda,\varepsilon) u_0 +
v_1(\lambda,\varepsilon), \\
&u^\lambda_2 = -\sigma_2(\lambda,\varepsilon) u_0 +
v_2(\lambda,\varepsilon), \\
&v_i(\lambda,\varepsilon) = o(|\sigma_i(\lambda,\varepsilon)|), \,\,
i=1,2.
\end{aligned} \right.
\end{equation}
Here $u_0 = (u_z,u_r,u_\theta)$ is the eigenfunction given by (\ref{eq5.6}), and
\begin{equation}
\label{eq5.12}
\left\{ \begin{aligned}
&\sigma_{1,2} = \frac{\sqrt{b^2(\varepsilon)+
4\beta^\varepsilon_1(\lambda)} \mp |b(\varepsilon)|}{2C} +
o(|b|,|\beta^\varepsilon_1|), \\
&\beta^\varepsilon_1(\lambda) = \alpha(\varepsilon)(\lambda -
\lambda^\varepsilon_0) + o(|\lambda-\lambda^\varepsilon_0|),
\,\,\alpha(0)>0, \\
&b(\varepsilon) = -\int_M (\widetilde w^\varepsilon \cdot \nabla
)w^\varepsilon \cdot w^{\varepsilon *}dx, 
\end{aligned} \right.
\end{equation}
where $w^\varepsilon =(\widetilde w^\varepsilon,w^\varepsilon_\theta)$,
with $\widetilde w^\varepsilon = (w^\varepsilon_z,w^\varepsilon_r)$, is
the eigenfunction of (\ref{eq5.2}) and (\ref{eq5.3}) corresponding
to $\lambda^\varepsilon_0$, $w^{\varepsilon *}$ is the dual
eigenfunction, and $C>0$ a constant.

\item The attractor $\Sigma^\varepsilon_\lambda$ attracts all bounded 
open sets of  $H \setminus \Gamma$, where
$\Gamma$ is the stable manifold of $u=0$ with co--dimension one. Namely, 
the space $H$ can be decomposed into two open sets
$U^\lambda_1$ and $U^\lambda_2$:
\[
H = \overline{U}^\lambda_1 + \overline{U}^\lambda_2, \qquad
U^\lambda_1 \cap \overline{U}^\lambda_2 = \emptyset, \qquad 0\in
\Gamma = \partial U^\lambda_1 \cap \partial U^\lambda_2,
\]
with $u^\lambda_i \in U^\lambda_i$ $(i=1,2)$, such that
\[
\lim_{t\to \infty} \|u(t,\varphi ) - u^\lambda_i \|_{L^2} =0, \quad
\text{when  $\varphi  \in U^\lambda_i$ $(i=1,2)$,}
\]
where $u(t,\varphi)$ is the solution of (\ref{eq2.18}) with
(\ref{eq2.15}) and (\ref{eq2.20}).
\end{enumerate}

\end{theorem}
\begin{remark}
{\rm
As mentioned in the Introduction, the existence of a bifurcation to a steady state solution is classical, and can be proved by the classical Krasnoselskii theorem. There are two  new ingredients in this theorem. First, here it is proved that the  the basin of attraction 
of the bifurcated attractor $\Sigma^\varepsilon_\lambda$ is $H \setminus \Gamma$. Second the explicit form of the bifurcated solutions. These new ingredients enable us to determine, in the next two sections,  the asymptotic structure in the physical space of the solutions of the problem, leading to justification of the flow structure of the  TCP flow.  }
\end{remark}

\begin{proof}[Proof of Theorem~\ref{th5.1}]
We shall apply Theorem~\ref{th4.6} to prove this theorem. Let $H$ and
$H_1$ be defined by (\ref{eq5.1}). We define the mappings
\begin{equation}
\label{eq5.13}
\left\{ \begin{aligned}
&L_\lambda =A + \lambda B:H_1 \to H, \\
&L^\varepsilon_\lambda = L_\lambda + \cL^\varepsilon_\lambda:H_1 \to
H, \\
&G:H_1 \to H,
\end{aligned} \right.
\end{equation}
by 
\begin{eqnarray*}
Au &=& \{P \Delta \widetilde u, \, \Delta u_\theta\}, \\
Bu &=& \{P(0,u_\theta), u_r\}, \\
\cL^\varepsilon_\lambda u &=&
\Bigg\{P\Bigg(\gamma(1-2\overline{r})u_r + \gamma(W_0 +
\overline{r}(1-\overline{r})) \frac{\partial
u_z}{\partial z}\, , \, \lambda(1-\mu)\overline{r} u_\theta \\
&&\quad + \,\gamma(W_0 + \overline{r}
(1-\overline{r}(1-\overline{r}))
\frac{\partial  u_r}{\partial z}\Bigg),\,
\gamma(W_0+\overline{r}(1-\overline{r}))\frac{\partial u_\theta}{\partial z}\Biggr\}, \\
G(u) &=& \left\{ -P(\widetilde u \cdot \nabla )\widetilde u,
-(\widetilde u\cdot \nabla )u_\theta\right\}.
\end{eqnarray*}
where $u= (u_z,u_r,u_\theta) \in H_1$, and the operator $P:L^2(M)^3
\to H$ is the Leray projection. Thus the equation (\ref{eq2.18})
with (\ref{eq2.15}) can be written in the abstract form
\[
\frac{du}{dt} =L_\lambda u + \cL^\varepsilon_\lambda u+G(u).
\]

It is well known that  the above defined operators satisfy  
the conditions (\ref{eq4.2})-(\ref{eq4.4}). In particular
\begin{align*}
& B : H_\alpha\longrightarrow H \,\,\text{is bounded for all $\alpha\ge
0$,} \\
& \cL^\varepsilon_\lambda :H_\alpha\longrightarrow H \,\,\text{is
bounded for all $\alpha\ge \frac{1}{2}$,} \\
& G : H_\alpha\longrightarrow H \,\,\text{is analytic for $\alpha >
\frac{1}{2}$,}
\end{align*}
where $H_\alpha$ is the fractional Sobolev spaces defined by the
interpolation between $H_0 =H$ and $H_1$.

It is clear that $L_\lambda$ is symmetric, and $G: H_\alpha \to H$
$(\alpha> \frac{1}{2})$ is bilinear satisfying that
$$\langle G(u,v),v\rangle_H =0$$
Hence, the condition (\ref{eq4.18}) holds true.

With the operators in (\ref{eq5.13}), the eigenvalue equation
(\ref{eq4.17}) corresponds to the problem (\ref{eq5.5}) with
(\ref{eq5.3}), and the eigenvector $e_0$ is precisely the  $u_0 =
(u_z,u_r,u_\theta)$ given by (\ref{eq5.6}).

Now consider the eigenvalue problem
\begin{equation}
\label{eq5.14}
L_\lambda u = \beta(\lambda)u, \qquad u = (u_z, u_r,u_\theta) \in
H_1.
\end{equation}
By (\ref{eq5.13}), the abstract form (\ref{eq5.14}) is equivalent  to
the following eigenvalue equations in $H_1$:
\begin{equation}
\label{eq5.15}
\left\{ \begin{aligned}
&\Delta u_z - \frac{\partial p}{\partial z} = \beta (\lambda) u_z, \\
&\Delta u_r - \frac{\partial p}{\partial r} + \lambda u_\theta =
\beta
(\lambda) u_r, \\
&\Delta u_\theta + \lambda u_r = \beta (\lambda) u_\theta, \\
& \divv\, \widetilde u =0.
\end{aligned} \right.
\end{equation}
It is known that the eigenvalues $\beta_k$ $(k=1,2,\cdots\,)$ of
(\ref{eq5.15}) in $H_1$ are real numbers satisfying
\begin{equation}
\label{eq5.16}
\left\{ \begin{aligned}
&\beta_1(\lambda) \ge \beta_2 (\lambda) \ge \cdots \ge
\beta_k(\lambda) \ge \cdots\, , \\
&\beta_k \longrightarrow -\infty \quad \text{as $k\to \infty$.}
\end{aligned}\right.
\end{equation}
Because the first eigenvalue $\lambda_0$ of (\ref{eq5.5}) with
(\ref{eq5.3}) is simple, the first eigenvalue $\beta_1(\lambda)$ of
(\ref{eq5.15}) with (\ref{eq5.3}) is also simple at $\lambda =
\lambda_0$. Noting that for the first eigenfunction $u_0$ of
(\ref{eq5.15}) with (\ref{eq5.3}) at $\lambda = \lambda_0$, which is
given by (\ref{eq5.6}), we have
\begin{eqnarray*}
\langle B u_0,u_0 \rangle_H &=& 2 \int_M u_r u_\theta dx \\
&=& \frac{2}{a^2 \lambda_0} \int^L_0 \cos^2 azdz \int^{r_1+1}_{r_1}
R \left( \frac{d^2}{dz^2} - a^2\right)^2 R dr \\ &=& \frac{2}{a^2
\lambda_0} \int^L_0 \cos^2 azdz \int^{r_1+1}_{r_1} \left|
\left( \frac{d^2}{dz^2} - a^2\right) R\right|^2  dr \\
&>& 0.
\end{eqnarray*}
Hence, from Theorem~\ref{th4.3} and (\ref{eq5.16}) we can derive the
conditions (\ref{eq4.5}) and (\ref{eq4.6}) at $\lambda=\lambda_0$.

Finally, by direct computation, we have  
$$G(u_0) \not= 0,$$ 
for the first eigenfunction $u_0$ of (\ref{eq5.5}) with (\ref{eq5.3}). By
Theorem~\ref{th4.9} one can derive that $u=0$ is a globally
asymptotically stable steady state solution of the following
equations with (\ref{eq5.3}) at $\lambda=\lambda_0$:
\begin{align*}
&\frac{\partial u_z}{\partial t} = \Delta u_z - \frac{\partial
p}{\partial z} - (\widetilde u \cdot \nabla ) u_z, \\
&\frac{\partial u_r}{\partial t} = \Delta u_r + \lambda u_\theta -
\frac{\partial p}{\partial r} - (\widetilde u \cdot \nabla ) u_r, \\
&\frac{\partial u_\theta}{\partial t} = \Delta u_\theta + \lambda u_r
- (\widetilde u\cdot \nabla ) u_\theta, \\
&\divv \, \widetilde u =0,
\end{align*} 
which correspond to the abstract equation (\ref{eq4.16}). Thus, this
theorem follows from Theorem~\ref{th4.6}. The proof is complete.

\section{Structural Transition of the Couette-Poiseuille 
Flow: Boundary-Layer Separation}
\label{sc3}
\setcounter{figure}{0}
In this and the next sections, we study the structure and its  transitions in  the CP and TCP flows, using a recently developed geometric theory of 2D incompressible flows by the authors; see a recent monograph by the authors  \cite{amsbook} and the references therein. The results obtained  
provide a rigorous characterization on how, when and where the Taylor vortices are generated.

\subsection{Geometric Theory of Incompressible Flows}
\label{geometry}
We first give a recapitulation of some results in the geometric theory of incompressible flows, which will be used in characterizing the structure and its transitions in the CP and TCP flows. We refer interested readers to above references for  further details of the theory.

\subsubsection{Structural stability}
\label{sc3.1}
Let $M\subset \R^2$ be an open set, and $C^r(M,\R^2)$ be the space of all
$C^r$ $(r\ge 1)$ vector field on $M$. We denote
\begin{eqnarray*}
D^r(M,\R^2) &=& \left\{ v\in C^r(M,\R^2)\mid \divv\, v =0, \,\, v\cdot
n|_{\partial \Omega} =0 \right\}, \\
B^r(M,\R^2) &=& \left\{ v\in D^r(M,\R^2) \mid \frac{\partial
v_n}{\partial \tau} =0 \,\,\text{on $\partial \Omega$}\right\}, \\
B^r_0 (M,\R^2) &=& \left\{v\in D^r (M,\R^2) \mid v =0 \,\,\text{on
$\partial \Omega$}\right\},
\end{eqnarray*}
where $\tau,n$ are the unit tangent and normal vectors on $\partial
M$.

Let $X = D^r(M,\R^2)$, or $B^r(M,\R^2)$, or $B^r_0 (M,\R^2)$ in the
following definitions.

\begin{definition}
\label{df3.1}
Two vector fields $u,v \in X$ are called topologically equivalent if
there exists a homeomorphism of $\varphi :M\to M$, which takes the
orbits of $u$ to orbits of $v$ and preserves their orientation.
\end{definition}

\begin{definition}
\label{df3.2}
A vector field $u\in X$ is called structurally stable in $X$ if there
exists a neighborhood $O\subset X$ of $u$ such that for any $v\in O$, $u$
and $v$ are topologically equivalent.
\end{definition}

For $u\in B^r_0(M,\R^2)$ $(r\ge 2$), a different singularity concept
for points on the boundary was introduced in \cite{mw01}, we proceed as
follows.

\begin{enumerate}
\item A point $p\in \partial M$ is called a $\partial $--regular
point of $u$ if
$$\frac{\partial u_\tau (p)}{\partial n} \not= 0,$$
and otherwise $p \in \partial M$ is called a $\partial $--singular
point.

\item A $\partial $--singular point $p \in \partial M$ of $u$ is
called nondegenerate if
$$\det \left( \begin{matrix}
\frac{\partial ^2u_\tau(p)}{\partial \tau\partial n} & \frac{\partial
^2 u_\tau(p)}{\partial n^2} \\
& \\
\frac{\partial ^2u_n(p)}{\partial \tau\partial n} & \frac{\partial ^2
u_n(p)}{\partial n^2}
\end{matrix} \right) \not= 0.$$
A nondegenerate $\partial $--singular point of $u$ is also called a
$\partial $--saddle point.

\item A vector field $u\in B^r_0(M,\R^2)$ $(r\ge 2)$ is called
$D$-regular if $u$ is regular in $M$, and all $\partial $--singular
points of $u$ on $\partial M$ are nondegenerate.
\end{enumerate}

The following theorem provides necessary and sufficient conditions
for structural stability in $B^r_0(M,\R^2)$. For the structural
stability theorems in $D^r(M,\R^2)$ and $B^r(M,\R^2)$, 
see \cite{amsbook,mw02a}.

\begin{theorem}[Ma and Wang \cite{mw01}]
\label{th3.3}
Let $u\in B^r_0(M,\R^2)$ $(r\ge 2)$. Then $u$ is structurally stable
in $B^r_0(M,\R^2)$ if and only if
\begin{enumerate}
\item $u$ is $D$-regular,

\item all interior saddle points of $u$ are self--connected; and

\item each $\partial $-saddle point of $u$ on $\partial M$ is
connected to a $\partial $-saddle point on the same connected
component of $\partial M$.
\end{enumerate}
Moreover, the set of all structurally stable vector fields is open
and dense in $B^r_0(M,\R^2)$.
\end{theorem}

\subsubsection{Structural bifurcation and boundary-layer separation}
\label{sc3.2}
Let $u(\cdot, \lambda) \in B^r_0(M,\R^2)$ be a one-parameter family
of vector fields. Assume that the boundary $\partial M$ contains a
flat part $\Gamma \subset \partial M$ and $x_0 \in \Gamma$.  For
simplicity, we take a coordinate system $(x_1,x_2)$ with $x_0$ at the
origin and with $\Gamma$ given by
$$\Gamma = \left\{ (x_1,0)\, \big| \, |x_1|<\delta \,\, \text{for some
$\delta>0$}\right\}.$$
Obviously, the tangent and normal vectors on $\Gamma$ are the unit
vectors in the $x_1$- and $x_2$-directions respectively. In a
neighborhood of $x_0 \in \Gamma$, $u(x,\lambda)$ can be expressed
near $x=0$ by
$$u(x,\lambda) = x_2v(x,\lambda).$$
To proceed, we consider the Taylor expansion of $u(x,\lambda)$ at
$\lambda=\lambda_0$:
\begin{equation}
\label{eq3.1}
\left\{ \begin{aligned}
&u(x,\lambda) = u^0(x) + (\lambda-\lambda_0) u^1(x) +
o(|\lambda-\lambda_0|), \\
&u^0(x) = x_2v^0(x).
\end{aligned} \right.
\end{equation}
Let $u^0 = (u^0_1, u^0_2)$. We assume that
\begin{align}
&
\label{eq3.2}
\frac{\partial u^0(0)}{\partial n} = v^0(0) =0,\\
&
\label{eq3.3}
\ind(v^0,0) =0,\\
&
\label{eq3.4}
\frac{\partial u^1(0)}{\partial n} \not= 0,,\\
&
\label{eq3.5}
\frac{\partial ^{k+1}u^0_1(0)}{\partial ^k\tau \partial n} \not= 0,
\quad \text{for some $k\ge 2$}.
\end{align}

The following is a theorem describing the structural bifurcation on
boundary for the case with index zero, which characterizes the
boundary-layer separation for 2-$D$ incompressible fluid flow. 
\begin{theorem}
\label{th3.4} Let $u(\cdot,\lambda) \in B^r_0(M,\R^2)$ be as given in
(\ref{eq3.1}) satisfying the conditions
(\ref{eq3.2})-(\ref{eq3.5}). Then the following assertions hold true:

\begin{enumerate}
\item As $\lambda<\lambda_0$ (or $\lambda>\lambda_0$), the flow
described by $u(x,\lambda)$ is topologically equivalent to a
parallel flow near $x_0\in \Gamma$, as shown in Figure~\ref{fg3.1}a.

\item As $\lambda_0<\lambda$ (or $\lambda>\lambda_0$), there are
some closed orbits of $u$ separated from $x_0 \in \Gamma$, i.e.\ some
vortices separated from $x_0 \in \Gamma$, as shown schematically in
either Figure~\ref{fg3.1} (c) or (d).

\item If $k=2$ in (\ref{eq3.5}), the vortex separated from $x_0 \in
\Gamma$ is unique, and the flow structure enjoys the following
properties:
\begin{enumerate}
\item there are exactly two $\partial$-saddle points $x^\pm =
(x^\pm_1,0) \in \Gamma$ of $u(x,\lambda)$ near $x_0 =0$ with
$x^-_1<0<x^+_1$, which are connected by an interior orbit
$\gamma(\lambda)$ of $u(x,\lambda)$;

\item the closed orbits separated from $x_0$ are enclosed by the
interior orbit $\gamma(\lambda)$ and the portion of the boundary
between $x^-$ and $x^+$; and

\item the interior orbit $\gamma(\lambda)$ shrinks to $x_0 $ as
$\lambda\to \lambda_0$.
\end{enumerate}
\end{enumerate}
\end{theorem}
\begin{figure}
 \centering \includegraphics[height=0.7\hsize]{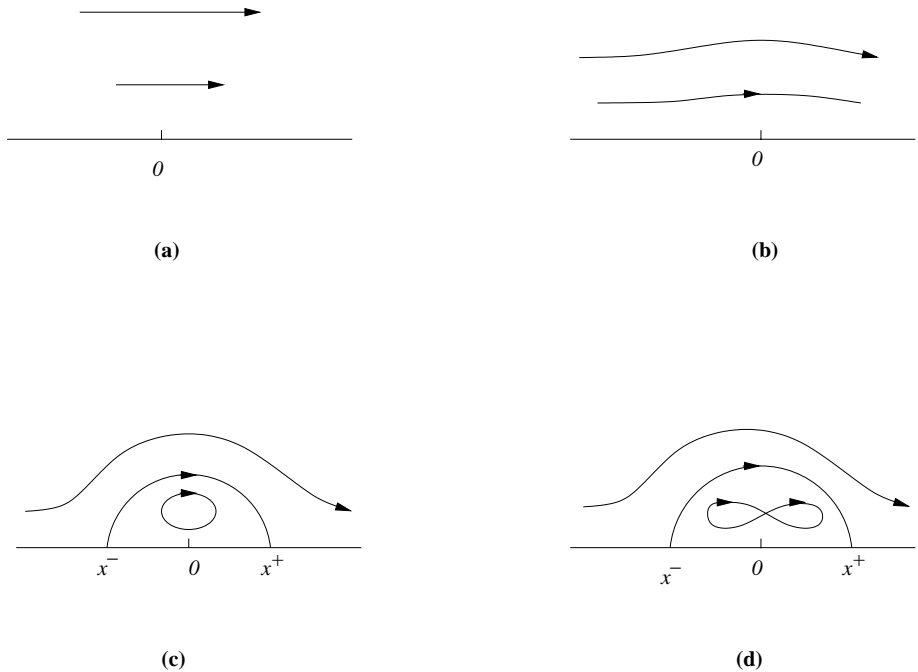}
\caption{Boundary-layer separation and re-attachment near 
a flat boundary.}
\label{fg3.1}
\end{figure}

We mention that when $\partial M$ is curved at $x_0 \in \partial M$, 
corresponding theorem is also true;  see \cite{amsbook,gmw1, gmw2}.

\subsection{Boundary-layer separation in the TCP flow}
\label{sc5.4} 
Hereafter, we always assume that the conditions in Theorem~\ref{th5.1} hold true. The main objective of this and the next sections is to study the asymptotic structure and its 
transition in the physical space in the CP flow and  the TCP flow regimes. 
As mentioned in the Introduction, the results obtained in these two sections provide a rigorous characterization on how, when and where the Taylor vortices are generated. In particular, contrary to what is commonly believed, we show that the 
propagating Taylor vortices (PTV) do not appear after the first dynamical bifurcation, and they appear only after further increase the Taylor number so that the boundary-layer or interior separation  occurs. 


We shall prove that the type of separations (boundary-layer or interior) is dictated by the (vertical) through flow driven by a pressure gradient along the rotating axis. Hence we consider two cases. The first is the case where through flow vanishes at the cylinder walls, i.e. $W_0 =0$ in (\ref{eq2.7}). This leads to boundary-layer separation, and will be studied in this section. 

The second is the case where $W_0\not=0$, leading to interior separations. This case will be addressed in the next section.

\subsubsection{Main results}
\label{sc5.4.1} 
Consider now the case where $W_0 =0$, and the main results are  the following two theorems. The first gives a precise  characterization on how, when and where the propagating Taylor vortices (PTV) are generated for the bifurcated solutions, and the second theorem describes time evolution to the PTV for the solutions of the equations.

\begin{theorem}
\label{th5.2}  Assume that the conditions in Theorem~\ref{th5.1} hold true, and assume that the constant velocity in boundary condition (\ref{eq2.7}) vanishes, i.e.,\ $W_0 =0$. Then there is a $\gamma_0>0$ such that 
for any $0<\gamma<\gamma_0$ where $\gamma$ is defined by (\ref{eq2.19}) , 
the following assertions hold true for  the two (bifurcated) steady state solutions
$v^\lambda_i =\mathcal U^{cp} + u^\lambda_i$ $(i=1,2)$ of (\ref{eq2.2}) and
(\ref{eq2.7}), where $\mathcal U^{cp}$ and $u^\lambda_i$ are given by (\ref{eq2.3})
and (\ref{eq5.11}):

\begin{enumerate}

\item For $v^\lambda_1$,  there is a number
$\lambda_1>\lambda^\varepsilon_0$ ($\lambda^\varepsilon_0$ as in
(\ref{eq5.4})) such that 

\begin{enumerate}

\item for  $\lambda^\varepsilon_0<\lambda<\lambda_1$,  
the vector field $\widetilde v^\lambda_1 = (W+u^\lambda_{1z}, u^\lambda_{1r})$  
is topologically equivalent to the vertical shear flow $(W,0)$ as shown 
in Figure~\ref{fg5.1}, and 

\item for $\lambda_1<\lambda$,  there is a unique center of $\widetilde
v^\lambda_1$ separated from each point $(z_n,r_1) \in \partial M$, where 
$z_n \simeq (4n+1)\pi/2 a$ (resp.  $(\bar z_n,r_2)\in \partial M$, $\bar z_n =
(4n+3)\pi/2a$) for $n=0,1,\cdots,k_0$, as shown in
Figure~\ref{fg5.2}, with $a$ as in (\ref{eq5.6}).

\end{enumerate}

\item For $v^\lambda_2$,  there is a number $b_0>0$ such that only one
of the following two assertions holds true:

\begin{enumerate}

\item If  $|b(\varepsilon)|<b_0$ $(b(\varepsilon)$ as in
(\ref{eq5.12})), there is a $\lambda_2>\lambda^\varepsilon_0$ such
that 

\begin{enumerate}

\item for  $\lambda^\varepsilon_0<\lambda<\lambda_2$,  the vector field
$\widetilde v^\lambda_2 = (W+u^\lambda_{2z}, u^\lambda_{2r})$  has the topological structure near $r=r_1$ (resp.  $r=r_2$) as shown in Figure~\ref{fg5.1}, and 

\item for $\lambda_2<\lambda$,  there is a unique center of $\widetilde
v^\lambda_2$ separated from each point $(\bar z_n,r_1)\in \partial M$ 
(resp.  $(z_n,r_2)\in \partial M)$ for $n=0,1,\cdots,k_0$, as shown in
Figure~\ref{fg5.2}.

\end{enumerate}

\item If $|b(\varepsilon)|>b_0$, then for  $\lambda>\lambda_2 =
\lambda^\varepsilon_0$, the vector field $\widetilde v^\lambda_2$ is
topologically equivalent to the structure as shown in
Figure~\ref{fg5.3}.
\end{enumerate}

\item There exists a $\lambda_3>\lambda_1$ $(\lambda_1\ge
\lambda_2)$ such that for  $\lambda_i<\lambda<\lambda_3$, the vector field
$\widetilde v^\lambda_i$ $(i=1,2)$ is topologically equivalent to
the structure as shown in Figure~\ref{fg5.3}, and $\lambda_3$ is
independent of $\gamma$.
\end{enumerate}
\end{theorem}

From Theorems \ref{th5.1}, \ref{th5.2} and \ref{th3.3}, 
we immediately derive the following
theorem, which links the dynamics with the structure in the physical space. 

\begin{theorem}
\label{th5.3} 
Assume that the conditions in Theorem~\ref{th5.1} hold true, 
$W_0 =0$ in (\ref{eq2.7}), and $0<\gamma<\gamma_0$. 
Then there are $\widetilde \lambda_j$ $(1\le
j\le 3)$ with $\widetilde \lambda_1 = \lambda_1$, $\lambda^* \le
\widetilde \lambda_2\le \lambda_2 $, $\widetilde \lambda_3 =
\lambda_3$, where $\lambda_j$ $(1\le j\le 3)$ are as in
Theorem~\ref{th5.2} and $\lambda^*$ is the saddle--node bifurcation
point (if it exists), otherwise $\lambda^* = \lambda^\varepsilon_0$,
and   for each $\lambda<\widetilde\lambda_3$ the space $H$ can be
decomposed into two open sets
$$H = \overline{U}^\lambda_1 + \overline{U}^\lambda_2, \quad
U^\lambda_1 \cap U^\lambda_2 = \emptyset,
$$
such that the following assertions hold true.

\begin{enumerate}
\item For $\varphi  \in U^\lambda_i$ there is a time $t_0 \ge 0$ such that if 
$\lambda<\widetilde \lambda_i$ $(i=1,2)$ for the solution $u(t,\varphi) =
(\widetilde u,u_\theta)$ of (\ref{eq2.2}) with (\ref{eq2.7}) and
(\ref{eq2.20}), the vector field $\widetilde u$ is topologically
equivalent to the structure as shown in Figure~\ref{fg5.1}, and if 
$\widetilde \lambda_i <\lambda<\widetilde \lambda_3$,  $\widetilde u$
is topologically equivalent to the structure as shown in
Figure~\ref{fg5.3} for all $t>t_0$.

\item In particular, for the initial value $\varphi \in U^\lambda_i$ near
the Couette--Poiseuille flow (\ref{eq2.3}), if  $\lambda>\widetilde
\lambda_i$ there is a time $t_0>0$ such that the solution $u(t,\varphi)$
of (\ref{eq2.2}) with (\ref{eq2.7}) and (\ref{eq2.20}) has a
boundary-layer separation at $t=t_0$, i.e.,\ the vector field
$\widetilde u$ of $u(t, \varphi)$ is topologically equivalent to the
structure as shown in Figure~\ref{fg5.1} for $t<t_0$, and
$\widetilde u$ is topologically equivalent to the structure near
boundary $\partial M$ as shown in Figure~\ref{fg5.2} for $t_0<t$.
\end{enumerate}
\end{theorem}
\begin{figure}
 \centering \includegraphics[height=0.45\hsize]{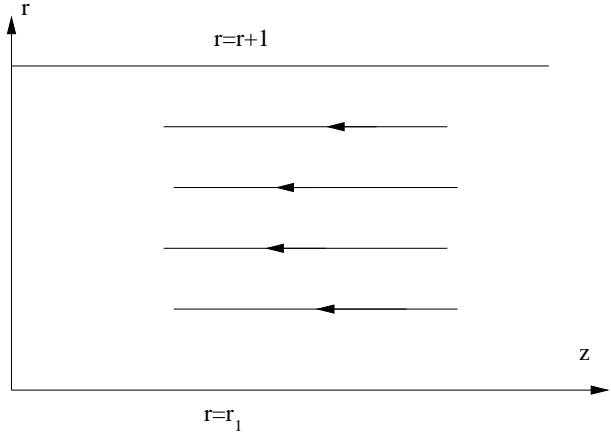}
\caption{Vertical shear flow with vanishing boundary velocity}
\label{fg5.1}
\end{figure}

\begin{figure}
 \centering \includegraphics[height=0.3\hsize]{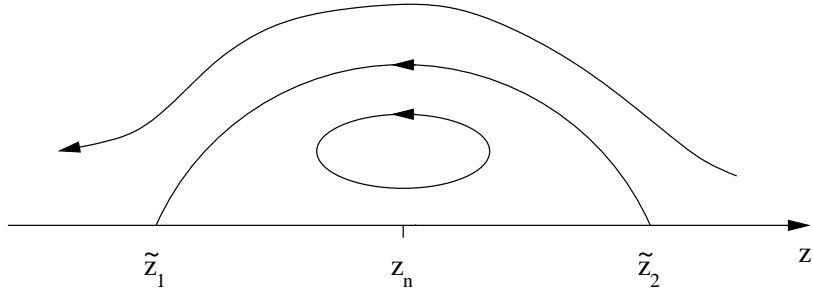}
\caption{Boundary separation and re-attachment.}
\label{fg5.2}
\end{figure}

\begin{figure}
 \centering \includegraphics[height=0.5\hsize]{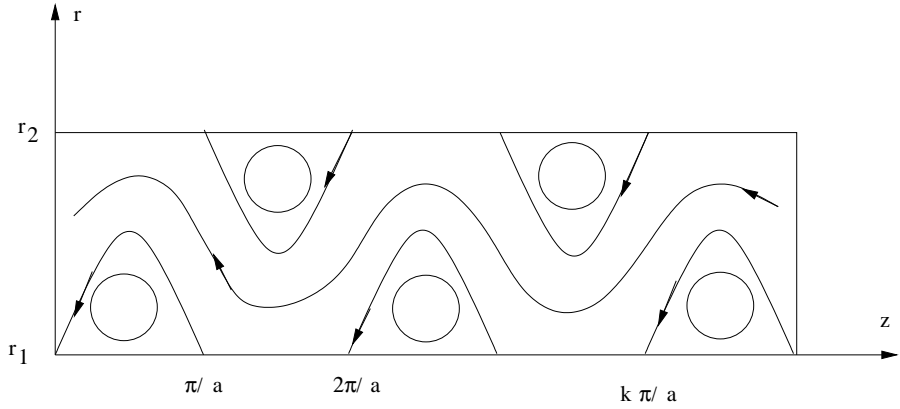}
\caption{Propagating Taylor Vortices after boundary-layer separations.}
\label{fg5.3}
\end{figure}

\begin{remark}
\label{rm5.4} 
{\rm
The critical values $\widetilde \lambda_i$ $(i=1,2)$
in Theorem~\ref{th5.3} which determine the boundary-layer separation
for the Taylor-Couette-Poiseuille flow problem depend  continuously 
on the non-dimensional parameter 
$\gamma = p_0d^3/4\rho\nu^2$. The expansion is as follows
\begin{equation}
\label{eq5.17}
\widetilde \lambda_i(\gamma) = \alpha_i \gamma + o(|\gamma|), \qquad
\alpha_i >0 \quad \text{a constant, $i=1,2$.}
\end{equation}
In addition, the condition $0 < \gamma < \gamma_0$ in the above theorems is the consequence of the narrow-gap assumption (i.e. $d$ is small).
}
\end{remark}

\begin{remark}
\label{rm5.5}
{\rm 
The height $L$ has an integer number times the length of $\pi/a$,
i.e.,\ $L=K \cdot \frac{\pi}{a}$ for some integer $K$. Hence the
number of vortices separated from $r=r_1$ and $r=r_2$ in
Theorems~\ref{th5.2} and \ref{th5.3} satisfies
\[
k_0 = \left\{ \begin{aligned}
& \left[ \frac{K+1}{2}\right] &&\text{ for $r=r_1$ (or $r=r_2$),} \\
& \left[ \frac{K}{2}\right]  &&\text{ for $r=r_2$ (or $r=r_1$).}
\end{aligned} \right.
\]
where $[\alpha] $ is the integer part of $n$.
}
\end{remark}

\subsubsection{Proof of Theorem~\ref{th5.2}}
\label{sc5.4.2}
We proceed in a few steps as follows.

 \medskip
 
{\sc Step 1.} By Theorem~\ref{th5.1}, as the Taylor number $T=\lambda^2$ satisfies
$T_c<T$ $(T_c =\lambda^{\varepsilon^2}_0)$, or equivalently
$\lambda^\varepsilon_0<\lambda$, the equations (\ref{eq2.2}) with
(\ref{eq2.7}) and (\ref{eq2.14}) generate from the basic flow
(\ref{eq2.3}) two steady state solutions
\[
v^\lambda_i = \left( W(r) + u^\lambda_{iz}, u^\lambda_{ir},\, V(r) +
u^\lambda_{i\theta}\right),\quad i=1,2,
\]
which are asymptotically stable under the axisymmetric perturbation.

It is clear that the number $b(\varepsilon)$ in (\ref{eq5.12})
satisfies
\[
\lim_{\varepsilon\to0} b(\varepsilon) =0, \qquad (\varepsilon=
(1-\mu,\gamma)).
\]
Hence, to prove this theorem, it suffices to proceed only for the
following two vector fields:
\begin{align}
&
\label{eq5.18} \widetilde V^\lambda_1 = (W(r),0) +
\frac{\sqrt{b^2+4\alpha(\lambda-\lambda^\varepsilon_0)} - b}{2C}\,
\, \widetilde u_0, \\
& 
\label{eq5.19}
\widetilde V^\lambda_2 = (W(r),0) + \frac{\sqrt{b^2 +
4\alpha(\lambda-\lambda^\varepsilon_0)} +b}{2C}\,\, \widetilde u_0,
\end{align}
where $b=b(\varepsilon)$, $\alpha = \alpha(\varepsilon)>0$ near
$\varepsilon=0$, $C>0$ a constant, 
\begin{eqnarray*}
&&W(r) = -\gamma \overline{r}(1-\overline{r}), \quad (\overline{r} =
r-r_1), \\
&&\widetilde u_0 = \left( \frac{1}{a}\, \sin \, azR'(r), -\cos
azR(r)\right), \end{eqnarray*}
and $R(r)$ is given by (\ref{eq5.7}).

\medskip

{\sc Step 2.} {\it Claim: $\widetilde V^\lambda_1$ has a boundary-layer separation on
$\lambda_1<\lambda$ for some $\lambda_1>\lambda^\varepsilon_0$.}

We shall apply the structural bifurcation theorem, Theorem~\ref{th3.4},  to prove this claim. To this end, let 
\[
\Lambda = \left(\sqrt{b^2 +4\alpha(\lambda-\lambda^\varepsilon_0)}
-b\right)/2aC, 
\]
which is an increasing function of
$\lambda$ with $\Lambda=0$ at $\lambda=\lambda^\varepsilon_0$. Thus
the vector field (\ref{eq5.18}) can be rewritten as
\begin{equation}
\label{eq5.20}
\widetilde V^\lambda_1 = \left(\Lambda \sin\, azR'(r) - \gamma
\overline{r}(1-\overline{r}), -a\Lambda \cos\, az R(r)\right).
\end{equation}

We discuss the structural bifurcation of (\ref{eq5.20}) only on the
portion $r=r_1$ of $\partial M$, and on this portion $r=r_1+1$ of
$\partial M$, the proof is the same. 

Since $R(r_1) = R'(r_1)=0$, the zero points of $\frac{\partial \widetilde
V^\lambda_1}{\partial n} = \frac{\partial \widetilde
V^\lambda_1}{\partial r}$ on $r=r_1$ are given by the solutions of
the equation
\begin{equation}
\label{eq5.21}
\Lambda \sin \, az R'' (r_1) -\gamma =0.
\end{equation}
Let $\Lambda_0 = \gamma/R''(r_1)$. Then, we infer from (\ref{eq5.21})
that   for $r=r_1$, 
\begin{equation}
\label{eq5.22}
\frac{\partial \widetilde V^\lambda_1}{\partial n} \left\{
\begin{aligned}
&\not= 0 \quad && \text{when } \Lambda<\Lambda_0, \\
&=0\quad &&\text{when } \Lambda=\Lambda_0 \text{  and  } z_n =
\frac{(4n+1)\pi}{2a}.
\end{aligned} \right.
\end{equation}

As in (\ref{eq3.1}), $\widetilde V^\lambda_1$ can be Taylor expanded  at
$\Lambda=\Lambda_0$ as follows:
\begin{equation}
\label{eq5.23}
\left\{ \begin{aligned}
&\widetilde V^\lambda_1 = V^0 +(\Lambda-\Lambda_0)V^1, \\
&V^0 = (V^0_z,V^0_r) = (\Lambda_0 \sin az
R'-\gamma\overline{r}(1-\overline{r}), -a\Lambda_0 \cos azR), \\
&V^1 = (V^1_z,V^1_r) = (\sin az R', -a \cos azR).
\end{aligned} \right.
\end{equation}
We derive from (\ref{eq5.7}) that 
\begin{align}
\label{eq5.24}
R''(r) =& -\alpha^2_0 \cos \alpha_0x \\
& \qquad + [\beta_1(\alpha^2_2 -
\alpha^2_1) + 2\beta_2\alpha_1\alpha_2] \cosh(\alpha_1x) \cos (\alpha_2
x ) \nonumber \\
&\qquad  +[\beta_2(\alpha^2_1-\alpha^2_2) + 2\beta_1\alpha_1\alpha_2] 
\sinh(\alpha_1x) \sin( \alpha_2x), \nonumber 
\end{align}
where $x = \overline{r} - \frac{1}{2}$. Inserting (\ref{eq5.8}) into
(\ref{eq5.24}) we obtain
\begin{equation}
\label{eq5.25}
R''(r_1) \simeq 8 + 1.7 e^{\frac{5}{2}} \simeq 28.
\end{equation}
Thus  we derive from (\ref{eq5.22}), (\ref{eq5.23}) and (\ref{eq5.25}) that
\begin{align*}
&\frac{\partial V^0}{\partial n} = \frac{\partial  V^0}{\partial r} =0
\quad && \text{at}\,\, (z_n,r_1) = \left( \frac{4n+1}{2a}\, \pi,
r_1\right), \\
&\frac{\partial V^1}{\partial n} = \frac{\partial V^1}{\partial r}
\simeq (28,0) \quad &&  \text{at $(z_n,r_1)$,} \\
&\frac{\partial^3V^0_z}{\partial \tau^2\partial n} = \frac{\partial
^3V^0_z}{\partial z^2\partial r} \simeq -28a^2\Lambda_0 \quad
&& \text{at $(z_n,r_1)$.}
\end{align*}
Namely,  the conditions (\ref{eq3.2}), (\ref{eq3.4}) and
(\ref{eq3.5}) are valid.

Now, we verify the condition (\ref{eq3.3}),
i.e.,\ we need to prove that
\begin{equation}
\label{eq5.26}
\ind\left( \frac{\partial \widetilde V^{\lambda_0}_1}{\partial
r},(z_n,r_1)\right) =0.
\end{equation}
To this end, we consider the equation
\begin{equation}
\label{eq5.27}
\frac{\partial V^\lambda_z}{\partial r} = \Lambda \sin az R''(r) -
\gamma(1-2\overline{r})=0.
\end{equation}
The function $R''(r)$ has the Taylor expansion at $r=r_1$ as follows
\begin{equation}
\label{eq5.28}
R''(r) = R''(r_1) + R''(r_1)\overline{r} + o(|\overline{r}|).
\end{equation}
From (\ref{eq5.7}) and (\ref{eq5.3}) we can derive that
\begin{equation}
\label{eq5.29}
R'''(r_1) \simeq -64 \times \frac{1.7}{2} - 10 \times e^{\frac{5}{2}}
\simeq -175.
\end{equation}
On the other hand, we note that
\[
\Lambda = \frac{\gamma\Lambda}{R''(r_1)\Lambda_0} =
\frac{\gamma\Lambda}{28\Lambda_0}\, .
\]
Thus, by (\ref{eq5.25}), (\ref{eq5.28}) and (\ref{eq5.29}), the
equation (\ref{eq5.27}) near $(z,r) = (z_n,r_1)$ can be rewritten in 
the following form
\begin{equation}
\label{eq5.30}
\left( 1 - \frac{\Lambda}{\Lambda_0} \, \sin az\right) + \left(
\frac{175 \Lambda \sin az}{28\Lambda_0} -z\right) \overline{r} +
o(|\overline{r}|) =0.
\end{equation}
Obviously, there is a $\delta>0$ such that the equation
(\ref{eq5.30}) has no solutions in $0<r-r_1 = \overline{r}<\delta$,
$|z-z_n|<\delta$ and $\Lambda<\Lambda_0$.  Namely, in a neighborhood of
$(z_n,r_1)$, the vector field $\frac{\partial \widetilde
V^\lambda_1}{\partial r}$ has no singular points for
$\Lambda<\Lambda_0$. Hence, by the invariance of the index sum of singular
points in a domain, one can get (\ref{eq5.26}).

Let $\lambda_1$ be the number that $\Lambda(\lambda_1) = \Lambda_0$. It
is clear that $\lambda_1>\lambda^\varepsilon_0$. Thus, this claim
follows from Theorem~\ref{th3.4}.

\medskip

{\sc Step 2.} The vector field $\widetilde V^\lambda_1$ has no singular
points in $M$ for $\lambda^\varepsilon_0<\lambda<\lambda_1$, and has
only one singular point in each domain $(z_k,z_{k+1}) \times
(r_1,r_1+1)$, $z_k = \frac{k\pi}{a}$ $(0\le k\le K = \frac{aL}{\pi}
-1)$, for all $\lambda>\lambda_1$.

From (\ref{eq5.20}) we find that the singular points of $\widetilde
V^\lambda_1$ must be in the lines $\ell_k = \left\{\left(
\frac{k\pi}{a} + \frac{\pi}{2a}, r\right)\mid
r_1<r<r_1+1\right\}\subset M$, and
\[
\widetilde V^\lambda_{1r} =0, \quad \widetilde V^\lambda_{1k} =
(-1)^k\Lambda R'(r) - \gamma\overline{r}(1-\overline{r}), \quad
\text{on $\ell_k$.}
\]
Therefore, to prove this claim we only consider this equation:
\begin{equation}
\label{eq5.31}
(-1)^k \frac{R'(r)}{R''(r_1)} -
\frac{\Lambda_0}{\Lambda}\, \overline{r}(1-\overline{r}) =0, \quad
0<\overline{r}<1.
\end{equation}
It is known that $R'(r)$ is odd and $\overline{r}(1-\overline{r})$ is
an even function on the variable $x = \overline{r} - \frac{1}{2}$.
These functions are illustrated by Figure~\ref{fg5.4}.

\begin{figure}
 \centering \includegraphics[height=0.5\hsize]{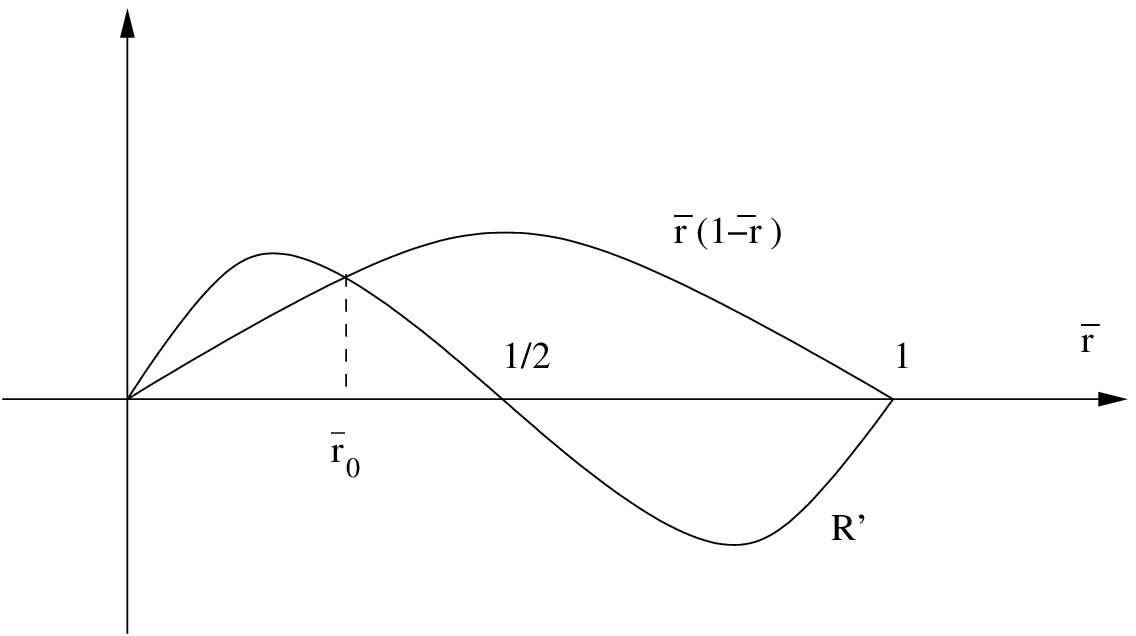}
\caption{$\overline r_0$ is a zero point of (\ref{eq5.31}).}
\label{fg5.4}
\end{figure}

Direct computations show that (\ref{eq5.31}) has no zero points in
$0<\overline{r}<1$ for $\Lambda \le \Lambda_0$, and there is a unique
zero point $\overline{r}_0 \in (0,1)$ for all $\Lambda_0<\Lambda$.
Furthermore, $0<\overline{r}_0 <\frac{1}{2}$ as $k=$ even, and
$\frac{1}{2} <\overline{r}_0<1$ as $k=$ odd. Thus, this claim is
proved.

{\sc Step 3.} By the invariance of index sum of singular points in a
domain, it is easy to see that the singular point of $\widetilde
V^\lambda_1$ in the domain $(z_k,z_{k+1})\times(r_1,r_2)$ must be a
center, which is enclosed by an orbit of $\widetilde V^\lambda_1$
connected to both boundary saddle points $\widetilde z_1$ and
$\widetilde z_2$, as shown in Figure~\ref{fg5.2}, $\widetilde z_1,
\widetilde z_2$ are on $r=r_1$ as $k=$ even, and on $r=r_2$ as $k=$
odd.

Finally, by Steps 1-3 one readily derives this theorem. The proof
is complete.

\section{Structural Transition of the Couette-Poiseuille 
Flow: Interior Separations}
In this section, we study the structure  and its transitions in the TCP flow 
for  the case where the $z$-direction
boundary velocity $W_0 \not= 0$ in (\ref{eq2.7}). 
We 
show that the secondary flow from the Couette--Poiseuille
flow (\ref{eq2.3}) will have  interior separations as the Taylor
number $T$ exceeds some critical value $\widetilde T$ which is an
increasing function of the nondimensional parameter  $\gamma = p_0d^3/4\rho\nu$.
For this purpose, we  recall some rigorous analysis by the authors 
on interior separations of incompressible flows; see \cite{mw04, amsbook}.

\subsection{Interior structural bifurcation of 2D incompressible Flows}
\label{sc3.3}

Following notations used in Section~\ref{geometry}, we let $u(\cdot,\lambda)\in D^r(M,\R^2)$ $(r\ge 1)$ have the Taylor
expansion at $\lambda=\lambda_0$ as follows
\begin{equation}
\label{eq3.6}
u(x,\lambda) = u^0(x) + (\lambda-\lambda_0)u^1(x) +
o(|\lambda-\lambda_0|).
\end{equation}

We assume that $x_0 \in M$ is an isolated singular point of $u_0(x)$,
and
\begin{align}
& 
\label{eq3.7}
\ind(u^0,x_0) =0,
\\ &
\label{eq3.8}
Du^0(x_0) = \left( \begin{matrix}
\frac{\partial u^0_1(x_0)}{\partial x_1} & \frac{\partial
u^0_1(x_0)}{\partial x_2} \\
& \\
\frac{\partial u^0_2(x_0)}{\partial x_1} &\frac{\partial
u^0_2(x_0)}{\partial x_2}
\end{matrix} \right) \not= 0,
\\&
\label{eq3.9}
u^1(x_0) \cdot e_2 \not=0,
\end{align}
where $e_2$ is a unit vector satisfying
\begin{equation}
\label{eq3.10}
\left\{ \begin{aligned}
&Du^0(x_0) e_2 = \alpha e_1 \qquad (\alpha\not= 0), \\
&Du^0(x_0) e_1 =0.
\end{aligned} \right.
\end{equation}
We also assume that $u^0\in C^m$ at $x_0 \in M$ for some even number
$m\ge 2$, and
\begin{equation}
\label{eq3.11}
\frac{\partial ^k(u^0(x_0) \cdot e_2)}{\partial e^k_1} = \left\{
\begin{aligned}
0,\qquad &1\le k<m = \text{even,} \\
\not= 0,\qquad & k=m=\text{even.}
\end{aligned} \right.
\end{equation}

Then we have the following interior structural bifurcation theorem,
which was proved in \cite{mw04,amsbook}.
\begin{theorem}
\label{th3.5}
Let $u(\cdot, \lambda) \in D^r(M,\R^2)$ be as given in (\ref{eq3.6})
satisfying the conditions (\ref{eq3.7})-(\ref{eq3.9}). Then the
following assertions hold true.
\begin{enumerate}
\item As $\lambda<\lambda_0$ (or $\lambda>\lambda_0$), the flow
described by $u(x,\lambda)$ is topologically equivalent to a tubular
flow near $x_0\in M$ as shown in Figure~\ref{fg3.2}(a).

\item As $\lambda>\lambda_0$ (or $\lambda<\lambda_0$) there must be
some centers of $u(x,\lambda)$ separated from $x_0\in M$ as shown
schematically in either Figure~\ref{fg3.2} (c) or (d).

\item The centers are enclosed by an extended orbit
$\gamma(\lambda)$, and $\gamma(\lambda)$ shrinks to $x_0$ as
$\lambda\to \lambda_0$.

\item If the condition (\ref{eq3.11}) is satisfied, then the center
separated from $x_0 \in M$ is unique, as shown in Figure~\ref{fg3.2}
(c).
\end{enumerate}
\end{theorem}
\begin{figure}
        \centering \includegraphics[height=0.8\hsize]{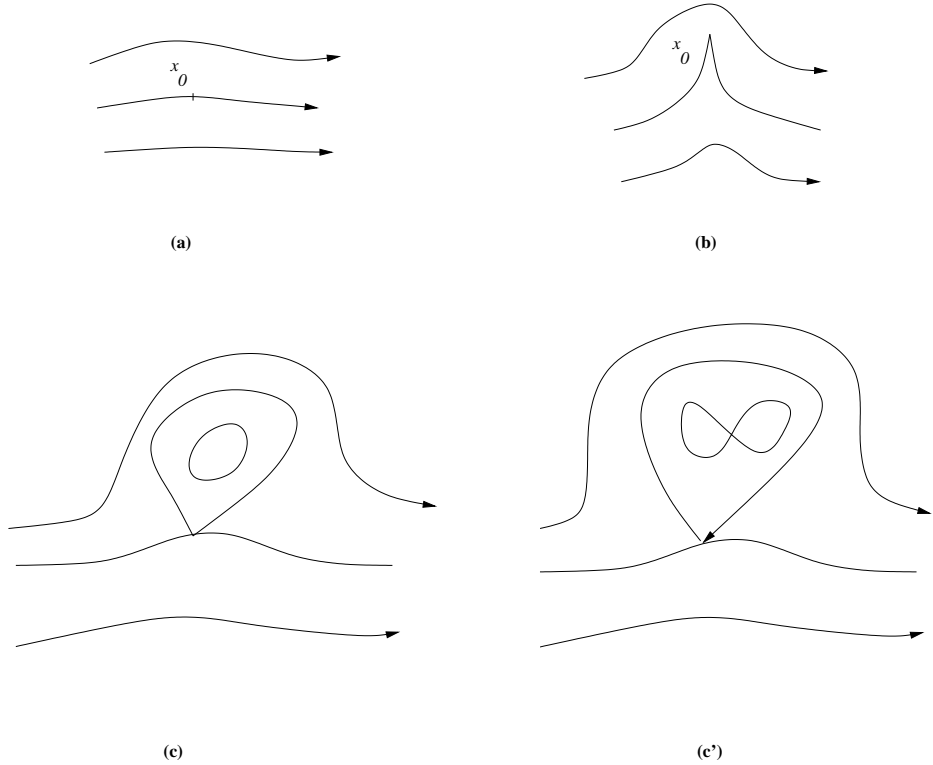}
        \caption{Interior separation.}
\label{fg3.2}
\end{figure}

\subsection{Interior separation of the Taylor-Couette-Poiseuille Flow}
\label{sc5.5}

\subsubsection{Main theorems}
\label{sc5.5.1}
Let the constant velocity $W_0 \not= 0$ in the boundary condition
(\ref{eq2.7}). Then, we have the following interior separation
theorems for the TCP flow problem. The results obtained here are in agreement 
with the numerical results obtained by Raguin and Georgiadis; 
see e.g. Figure~8 in \cite{rg}.

\begin{theorem}
\label{th5.6}
Assume that the conditions in Theorem~\ref{th5.1} hold true, and 
let $W_0 \not= 0$ in (\ref{eq2.7}). Then there exists
$\gamma_0>0$ such that if $0<\gamma<\gamma_0$,  then for the two (bifurcated) steady state solutions $v^\lambda_1 =\mathcal U^{cp} + u^\lambda_i$ $(i=1,2)$ of (\ref{eq2.2}) and
(\ref{eq2.7}), the following assertions hold true:

\begin{enumerate}

\item For $v^\lambda_1$,  there is a $\lambda_1 >\lambda^\varepsilon_0$
such that
\begin{enumerate}

\item for  $\lambda^\varepsilon_0<\lambda<\lambda_1$,  the vector
field $\widetilde v^\lambda_1 = (W+u^\lambda_{1z}, u^\lambda_{1r})$
 is topologically equivalent to the structure as
shown in Figure~\ref{fg5.5}, and 

\item for  $\lambda_1<\lambda$,  there is
exactly a pair of center and saddle points separated from a  point in
each domain $(\tilde z_k, \tilde z_{k+1}) \times (r_1,r_2) \subset M$, with $\tilde z_k =
\frac{k\pi}{a}$ $(0\le k\le K)$, as shown in Figure~\ref{fg3.2}(c).
Moreover, the value $\lambda_1$ is a continuous and increasing
function of $\gamma$.
\end{enumerate}

\item For $v^\lambda_2$ there is a number $b_0>0$ such that only one
of the following two assertions holds true.
\begin{enumerate}
\item If  $|b(\varepsilon)|<b_0$, where $b(\varepsilon)$ is given by 
(\ref{eq5.12})), then there is a $\lambda_2>\lambda^\varepsilon_0$
$(\lambda_1\ge \lambda_2)$ such that the same conclusions as  Assertion (1) 
holds true  for $v^\lambda_2$ with  $\lambda_2$  replacing $\lambda_1$.

\item If  $|b(\varepsilon)|>b_0$, then when  $\lambda>\lambda_2 =
\lambda^\varepsilon_0$,  the vector field $\widetilde v^\lambda_2 = (W
+ u^\lambda_{2z},u^\lambda_{2r})$ is topologically equivalent to the
structure as shown in Figure~\ref{fg5.6}.
\end{enumerate}

\item There exists a $\lambda_3 >\lambda_1$,  with $\lambda_3$
independent of $\gamma$, such that  if  $\lambda_i<\lambda<\lambda_3$, 
the vector field $\widetilde v^\lambda_i$ $(i=1,2)$ is topologically
equivalent to the structure as shown in Figure~\ref{fg5.6}.
\end{enumerate}
\end{theorem}

The following theorem is a direct corollary of  Theorems
\ref{th3.3}, \ref{th5.1} and \ref{th5.6}, and provides a link between the dynamics and the structure in the physical space.

\begin{theorem}
\label{th5.7}
Assume that the conditions in Theorem~\ref{th5.1} hold true, 
 $W_0 \not=0$ in (\ref{eq2.7}), and $0<\gamma<\gamma_0$. Then
there are $\widetilde \lambda_j$ $(1\le j\le 3)$ with $\widetilde
\lambda_1 = \lambda_1$, $\lambda^* \le \widetilde \lambda_2 \le
\lambda_2$, $\widetilde \lambda_3 = \lambda_3$, and for each
$\lambda<\widetilde \lambda_3$ the space $H$ can be decomposed into
two open sets
\[
H = \overline{U}^\lambda_1 + \overline{U}^\lambda_2, \qquad
U^\lambda_1 \cap U^\lambda_2 = \emptyset,
\]
such that the following assertions hold true.

\begin{enumerate}
\item For $\varphi  \in U^\lambda_i$ there is a time $t_0\ge 0$   such  that 
when  $t > t_0$, for the solution $u(t,  \varphi ) = (\widetilde u,u_\theta)$ of
(\ref{eq2.2}) with (\ref{eq2.7}) and (\ref{eq2.20}), as
$\lambda<\widetilde \lambda_i$ the vector field $\widetilde u$ is
topologically equivalent to the structure as shown in
Figure~\ref{fg5.5}, and as $\widetilde \lambda_i <\lambda<\widetilde
\lambda_3$ $(i=1,2)$ $\widetilde u$ is topologically equivalent to
the structure as shown in Figure~\ref{fg5.6}.

\item In particular, for the initial value $\varphi \in U^\lambda_i$ near
the Couette--Poiseuille flow (\ref{eq2.3}), and for  $\lambda>\widetilde
\lambda_i$ $(i=1,2)$,  there is a time $t_0>0$ such that for the
solution $u(t,\varphi) = (\widetilde u,u_\theta)$ of (\ref{eq2.2}) with
(\ref{eq2.7}) and (\ref{eq2.20}), $\widetilde u$ has an interior
separation at $t=t_0$ from a point in each domain $(\tilde z_k, \tilde z_{k+1})
\times (r_1,r_2) \subset M$, where $\tilde z_k = \frac{k\pi}{a}$, as described
by Theorem~\ref{th5.6} with $t$ replacing $\lambda$.
\end{enumerate}
\end{theorem}
\begin{figure}
        \centering \includegraphics[height=0.5\hsize]{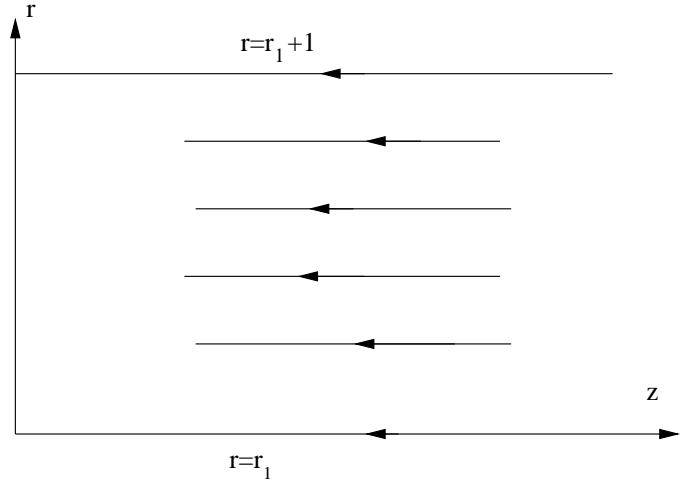}
        \caption{Vertical shear flow with a constant velocity $W_0 \not=0$ 
on boundary.}
\label{fg5.5}
\end{figure}
\begin{figure}
        \centering \includegraphics[height=0.35\hsize]{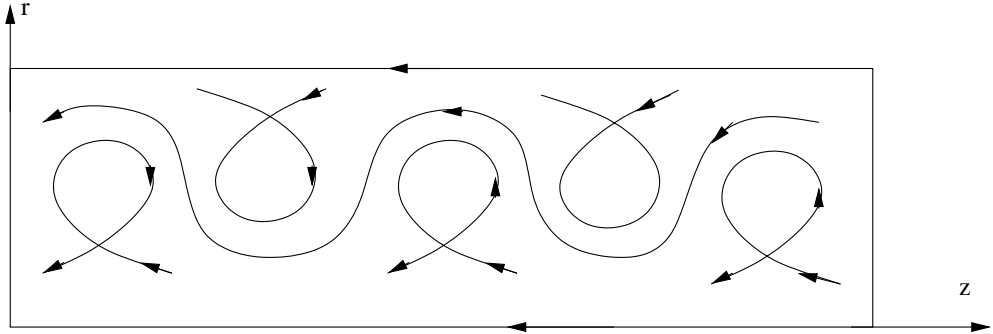}
        \caption{Propagating Taylor vortices separated from the interior.}
\label{fg5.6}
\end{figure}

\subsubsection{Proof of Theorem~\ref{th5.6}}
\label{sc5.5.2}
This proof is similar to that of Theorem~\ref{th5.2}. Here, we only
need to prove the interior structural bifurcation in Assertion~(1)
for the following vector field
\begin{equation}
\label{eq5.32}
\widetilde V^\lambda_1 = \left( \Lambda \sin azR'(r) -
\gamma\overline{r}(1-\overline{r}) - \gamma W_0, -a\Lambda\cos az
R(r)\right).
\end{equation}

We proceed by applying Theorem~\ref{th3.5}. Consider the equations
\begin{equation}
\label{eq5.33}
\left\{ \begin{aligned}
&\pm \Lambda R'(r) - \gamma\overline{r}(1-\overline{r}) - \gamma W_0
=0, \\
&\pm \Lambda R''(r) - \gamma(1-2\overline{r}) =0.
\end{aligned} \right.
\end{equation}
By computing, we can know that the equations (\ref{eq5.33}) have a
unique solution $(\Lambda_0,r^\pm_0)$ with $r^+_0 - r_1 = 1 -(r_0^- -
r_1)$ and $r^+_0 <r^+_*<r^-_* <r^-_0$, where $r^\pm_*$ are the
maximum points of $R'(r)$, i.e.
\begin{equation}
\label{eq5.34} R''(r^\pm_*) =0, \qquad \textstyle{\left(r^+_* \simeq
r_1 + \frac{1}{5},\, r^-_* \simeq r_1 + \frac{4}{5}\right)}.
\end{equation}
The vector field (\ref{eq5.32}) has the following expression at
$\Lambda=\Lambda_0$
\begin{equation}
\label{eq5.35}
\left\{ \begin{aligned}
&\widetilde V^\lambda_1 = V^0 + (\Lambda-\Lambda_0)V^1, \\
&V^0 = (V^0_z,V^0_r) = (\Lambda_0 \sin az R' - \gamma
\overline{r}(1-\overline{r}) - \gamma W_0, -a\Lambda_0 \cos azR), \\
&V^1 = (V^1_z,V^1_r) = (\sin az R', -a \cos azR).
\end{aligned}
\right.
\end{equation}
It is easy to see that for the solution $(\Lambda_0, r^\pm_0)$ of
(\ref{eq5.33}) the points $(z_{2k}, r^+_0)$ and $(z_{2k+1},r^-_0)\in
M$ are singular points of $V^0$, where $z_m = \frac{m\pi}{a} +
\frac{\pi}{2a}$.

For simplicity, we only consider the structural bifurcation of $V^0$
at the singular point $(z_0,r_0) = (\frac{\pi}{2a},r^+_0)$.

We can check that when $\Lambda<\Lambda_0$, the vector field
$\widetilde V^\lambda_1$ given by (\ref{eq5.32}) has no singular
point in $M$, and $(z_0,r_0)$ is an isolated singular point of
$\widetilde V^\lambda_1 $ at $\Lambda=\Lambda_0$. Therefore we have
\begin{equation}
\label{eq5.36}
\ind(V^0,(z_0,r_0)) =0.
\end{equation}
In addition, we see that
\begin{equation}
\label{eq5.37}
DV^0 (z_0,r_0) = \left( \begin{matrix}
\frac{\partial V^0_z}{\partial z} & \frac{\partial V^0_z}{\partial r}
\\
& \\
\frac{\partial V^0_r}{\partial z} &\frac{\partial V^0_r}{\partial r}
\end{matrix} \right)_{(z_0,r_0)} = \left( \begin{matrix}
0 & 0 \quad\\
& \\
a^2\Lambda R(r_0) & 0
\end{matrix} \right) \not= 0
\end{equation}
and the eigenvectors $e_1$ and $e_2$ satisfying
\[
\left\{ \begin{aligned}
&DV^0 (z_0,r_0) e_2 = \alpha e_1, \qquad (\alpha=a^2\Lambda R(r_0)), \\
&DV^0 (z_0,r_0) e_1 =0,
\end{aligned} \right.
\]
are given by
\[
e_1 = (0,1), \qquad e_2 = (1,0).
\]
Hence, we have
\begin{equation}
\label{eq5.38}
V^1(z_0,r_0)\cdot e_2 = V^1_z (z_0,r_0) = R'(r_0) \not= 0.
\end{equation}
From (\ref{eq5.36})--(\ref{eq5.38}) we find that the conditions
(\ref{eq3.7})--(\ref{eq3.9}) are satisfied by the vector field
$\widetilde V^\lambda_1$ at $\Lambda=\Lambda_0$ and $(z,r) =
(z_0,r_0)$.

Finally, we verify the condition (\ref{eq3.11}). By (\ref{eq5.33}) we
have
\begin{equation}
\label{eq5.39}
\frac{\partial ^k(V^0\cdot e_2)}{\partial e^k_1}\bigg|_{(z_0,r_0)} =
\frac{\partial ^kV^0_z(z_0,r_0)}{\partial r^k} =0 \quad \text{for
$k=0,1$,}
\end{equation}
and
\begin{equation}
\label{eq5.40}
\frac{\partial ^2V^0_z(z_0,r_0)}{\partial r^2} = \Lambda_0 R'''(r_0) +
2\gamma.
\end{equation}
By (\ref{eq5.33}) we see that
\begin{equation}
\label{eq5.41}
\frac{\gamma}{\Lambda_0} = \frac{R''(r)}{1-2(r_0-r_1)}\, .
\end{equation}
On the other hand, we can check that
\begin{equation}
\label{eq5.42}
- \frac{R'''(r)}{R''(r)} > \frac{2}{1-2(r-r_1)}\,, \quad \forall\,\,
r_1\le r<r^+_*,
\end{equation}
where $r^+_*$ satisfies (\ref{eq5.34}). From
(\ref{eq5.40})--(\ref{eq5.42}) it follows that
\begin{equation}
\label{eq5.43}
\frac{\partial ^2V^0_z(z_0,r_0)}{\partial r^2} \not= 0.
\end{equation}
Thus, from (\ref{eq5.39}) and (\ref{eq5.43}) we derive the condition
(\ref{eq3.11}) with $m=2$. By Theorem~\ref{th3.5}, the vector field
$\widetilde V^\lambda_1$ has an interior separation from each point
$(y_m,r_0)$ at $\Lambda=\Lambda_0$, where $r_0 =r^+_0$ if $y_m =
\frac{2k\pi}{a} + \frac{\pi}{2a}$ and $r_0 = r^-_0$ if $y_m =
\frac{(2k+1)}{a}\, \pi + \frac{\pi}{2a}$. The proof is complete.
\end{proof}

\bibliographystyle{siam}

\bibliography{tcp}

\end{document}